\documentclass[%
twocolumn,
nofootinbib,
 amsmath,amssymb,
 aps,
pra,
]{revtex4-2}
\usepackage{graphicx}
\usepackage{dcolumn}
\usepackage{bm}

\usepackage{color,soul}
\usepackage{xcolor,hyperref}
\hypersetup{
    colorlinks=true,
    linkcolor=[rgb]{0 0 0.5},
    urlcolor=[rgb]{0 0 0.5},
    citecolor= [rgb]{0 0 0.5}
}

\newcommand{\comment}[1]{}

\newcommand{\la}[0]{\left\langle}
\newcommand{\ra}[0]{\right\rangle}
\renewcommand{\ol}[0]{\overline} 

\renewcommand{\d}{{\rm d}}

\newcommand{\BEA}{\begin{eqnarray}}
\newcommand{\EEA}{\end{eqnarray}}

\newcommand{\BvL}{Bohr-Van Leeuwen~}
\newcommand{\R}{{\bf R}}
\newcommand{\V}{{\bf V}}
\newcommand{\A}{{\bf A}}
\renewcommand{\r}{{\bf r}}
\renewcommand{\v}{{\bf v}}

\newcommand{\ez}{{\bf e}_z}
\newcommand{\bxi}{\boldsymbol{\xi}}

\begin{document}

\title{Lasting effects of static magnetic field on classical Brownian motion}

\author{\vspace{-0.5ex}Ashot Matevosyan$^{1,2)}$ and Armen E. Allahverdyan$^{2)}$ }

\affiliation{$^{1)}$University of Cambridge, Cavendish Laboratory, 19 J.J. Thompson Avenue, Cambridge CB3 0HE, UK,\\
$^{2)}$Alikhanyan National
  Laboratory (Yerevan Physics Institute), \\ Alikhanian Brothers Street 2, Yerevan 375036, Armenia}

\begin{abstract}
The \BvL theorem states that an external static magnetic field does not influence the state of a classical equilibrium system: there is no equilibrium classical magnetism, since the magnetic field does not do work. We revisit this famous no-go result and consider a classical charged Brownian particle interacting with an equilibrium bath. We confirm that the \BvL theorem holds for the long-time (equilibrium) state of the particle. But the external static, homogeneous magnetic field does influence the long-time state of the thermal bath, which is described via the Caldeira-Leggett model. In particular, the magnetic field induces an average angular momentum for the (uncharged!) bath, which separates into two sets rotating in opposite directions. The effect relates to the bath going slightly out of equilibrium under the influence of the Brownian particle and persists for arbitrarily long times. In this context we studied the behavior of the two other additive integrals of motion, energy and linear momentum. The situation with linear momentum is different, because it is dissipated away by (and from) the bath modes. The average energy of the bath mode retains the magnetic field as a small correction. Thus, only the bath angular momentum really feels the magnetic field for long times.

\comment{Justification

In classical physics, the equilibrium effects of static magnetic fields are forbidden by the famous Bohr-Van Leeuwen theorem. In this letter, we present a new view on this theorem: we show that long-time effects of the magnetic field can be found in the thermal bath that ensures the equilibration of the system. This result is important for various ﬁelds including plasma and fusion research, colloids, and biophysics, where magnetic fields are relevant. Our findings are demonstrated via two complementary approaches: the microscopic Caldeira-Leggett model and a coarse-grained hydrodynamic setup. 
}

\end{abstract}

\maketitle



\section{Introduction}

The \BvL theorem governs the response of equilibrium systems to external magnetic fields \cite{vleck,kaplan,seifert}. It states that charged classical particles do not feel the magnetic field in equilibrium. Thus, equilibrium influences of the magnetic field are to be restricted to the quantum domain, where the theorem does not apply due to the non-commutativity of quantum observables \cite{landau}. The essence of the theorem is that the equilibrium state can be represented as a function of the energy; e.g. the Gibbs  coordinate-velocity distribution, or the microcanonical distribution. Since the magnetic field does not do work, it does not appear in the energy that enters equilibrium distributions \cite{vleck,landau}; see section \ref{sec-bvl} for a reminder on this theorem and its origin. 

The \BvL theorem prevents the existence of equilibrium classical magnetism \cite{vleck}, a subject that could have potential applications in various fields including real and complex plasma \cite{weibel,book_plasma_colloid,plasma_rotation}, where the influence of magnetic fields is relevant for the fusion research, charged colloidal liquids \cite{book_plasma_colloid,lowen}, macroions \cite{aronson} {\it etc}. Especially interesting are biophysical applications\cite{zarema,albu,adair,glaser,bialek}: since a stationary magnetic field is not screened by a living body, it makes an interesting diagnostic tool and also a potential abusive factor for bio-systems. Moreover, there is a massive body of experimental results witnessing observable effects of weak, static magnetic fields in biological systems; see e.g. \cite{zarema,albu}. A natural target for the magnetic field is metal ions (Na$^+$, K$^+$, Ca$^{2+}$ {\it etc}) that are crucial in molecular biology: nearly 1/3 of all proteins employ metal ions for their functioning \cite{glaser}. Ions are important in bioenergetics, communication (e.g., nerve impulse generation), osmotic regulation, metabolism, energy storage {\it etc} \cite{bialek}. However, the translational motion of ions is always classical, hence the \BvL theorem prohibits their equilibrium magnetic response \cite{adair}. Biophysical responses to static magnetic fields might be looked for in the quantum domain; see e.g. \cite{briegel,cai,kominis}. It is however not likely that all biophysical influences of magnetic fields can be accounted for by quantum models \cite{adair}. 

In all these fields one deals with classical charges moving in the thermal baths. Thus, it is necessary to understand which effects are not prohibited by the \BvL theorem and can support long-time (i.e. lasting) influences of a static magnetic field for classical Brownian motion of a charged particle.    

Consider a Brownian charged particle described via the Langevin equation. The equilibrium state feels no magnetic field according to the \BvL theorem. This is however not the end of the story, because during its relaxation (whatever short) the particle perturbs the equilibrium state of the bath, which is now slightly out of equilibrium. Using the Caldeira-Leggett (CL) model \cite{magal,zwanzig,leggett,petr,tuckerman} that reproduces the Langevin dynamics, we show that an external static magnetic field leads to long-time changes in the bath state. The (uncharged) bath oscillators acquire a sizable average angular momentum and separate (in the frequency space) into two groups that rotate in different directions. This bath angular momentum can (but need not) be driven by the  conservation law of the total (for the bath + Brownian particle), effective angular momentum. Thus, the magnetic field can persist in the long-time limit, though its influence is to be looked for not in the state of the Brownian charge, but rather in its environment. 

Recall that CL is a concrete microscopic model composed of undamped harmonic oscillators. It has a wide range of applications for baths composed of weakly interacting particles (photons, phonons) \cite{leggett,petr}.  Its applicability to the real Brownian dynamics was clarified within molecular dynamic theories, where the oscillators refer to effective modes of a liquid bath \cite{molecular,stratt,tuckerman}. 

Our study of angular momentum led us to a more general physical question: how do the three additive integrals of motion (angular momentum, linear momentum and energy) behave at the interface between the system (Brownian particle) and bath? For example, is the bath capable of storing linear momentum in the same way it stores angular momentum? The answer to this question is negative: the bath angular momentum differs from the linear momentum and energy; see sections \ref{sec-energy} and \ref{sec-linear} for details. For the energy the situation is less interesting, since the influence of the Brownian particle amounts to a small perturbation of the bath energy. The linear momentum is more interesting, because it is transferred to the bath can be a dominant effect for finite times, but is eventually dissipated from its observables.

This paper is organized as follows. The next section discusses the CL model: an ion interacting with many independent harmonic oscillators representing an equilibrium thermal bath. Section \ref{sec-bvl} presents the famous \BvL theorem with discussion. In Section \ref{sec-angular} and \ref{sec-energy} we study the angular momentum  and total energy of individual modes respectively. We compare the behaviour of angular momentum with the linear momentum in the section \ref{sec-linear}. We summarize in the last section and provide a perspective on future research. We relegated detailed derivations of our results to appendixes making our conclusions self-contained. 
Appendix \ref{app-correlations} discusses the solution of the Langevin equation (\ref{lang}) and studies pertinent correlation functions of the Brownian motion. Appendix \ref{app-langevin} discusses the system-bath (Caldeira-Leggett) model and the derivation of the Langevin equation. in Appendix \ref{app-angular-momentum} calculated the mean angular momentum of a single bath oscillator, while Appendix \ref{app-energy} studies its mean energy.

\section{The Model} 
\label{sec-cl}

Consider a classical particle with coordinates $\R=(X,Y,Z)$, unit charge and unit mass that interacts with magnetic field. The particle is subject to an external, rotation symmetric, harmonic potential with frequency $\omega_0$. Particle's Lagrangian reads
\BEA
\label{lagr_sys}
\mathcal{L}_{S}=\frac{1}{2}\dot{\R}^2 -\frac{\omega_0^2}{2}\R^2+\A(\R)\dot\R,
\EEA
where $A(\R)$ is a vector potential that generates a static, homogeneous magnetic field
$\bm{B}$ with the magnitude $b=|\bm{B}|$ along the $z$-axes (the normal vector $\ez$):
\BEA
\label{mag}
\bm{B}={\rm rot}\,\A=\ez b, \quad \A(\R)
=\frac{1}{2}(-{bY}, {bX}, 0).
\EEA
The particle couples with a bath made of $N$ harmonic oscillators (modes) with coordinates $\r_k=(x_k,y_k,z_k)$, masses $m_k$, frequencies $\omega_k$ and coupling constants $c_k$. 
The potential energy of particle-bath interaction is assumed to be non-negative and bi-linear over the particle and bath coordinates (Caldeira-Leggett model). Hence the bath\,\texttt{+}\,interaction Lagrangian reads \cite{magal,petr,zwanzig,leggett,tuckerman}
\BEA
\label{lagr_bath}
\mathcal{L}_{B}={\sum}_{k=1}^N \left[\frac{m_k}{2}\dot{\r}_k^2 -
\frac{m_{k}\omega_{k}^{2}}{2} \left(\r_{k}-\frac{c_k\R}{m_k\omega_k^2}\right)^{2}\right],
\EEA
where the full Lagrangian is $\mathcal{L}_{S}+\mathcal{L}_{B}$. Due to (\ref{mag}) and linearity of equations of motion generated by $\mathcal{L}_{S}+\mathcal{L}_{B}$, the motion along $z$ coordinates for all particles involved in (\ref{lagr_sys}, \ref{lagr_bath}) decouples from the motion along $(x,y)$ coordinates. 

\comment{
We can direct the magnetic field $\bm{B}$ with the magnitude $b=|\bm{B}|$ along the $z$-axes (the normal vector $\ez$):
\BEA
\label{mag}
\bm{B}={\rm rot}\A=\ez b, ~~ \A=(A_x,A_y,A_z)=(-\frac{bY}{2}, \frac{bX}{2}, 0).
\EEA
We focus on the former motion described by 
\BEA
\label{pa}
\R\equiv(X,Y),\qquad \r_k=(x_k,y_k),
\EEA
and write for the vector-potential [cf.~(\ref{mag})]: 
\BEA
\A=(A_x,A_y)=(-\frac{bY}{2}, \frac{bX}{2}).
\EEA
The full Lagrangian $\mathcal{L}_{S}+\mathcal{L}_{B}$ is invariant (up to a total time-derivative) under rotations along z-axis: 
\BEA
\bm{\phi}=
\begin{pmatrix}
\phi_x\\
\phi_y
\end{pmatrix}\to
\begin{pmatrix}
\cos\varphi & \sin\varphi\\
-\sin\varphi & \cos\varphi
\end{pmatrix}
\begin{pmatrix}
\phi_x\\
\phi_y
\end{pmatrix}, 
\label{gomesh}
\EEA
where $\bm{\phi}=\R,\,\r_k,\,\bm{A}$. 
Taking in (\ref{gomesh}) infinitesimal ($\varphi\ll 1$) rotations, we get from Noether's theorem|or directly
}

We get from Noether's theorem, or directly from equations of motion generated by $\mathcal{L}_{S}+\mathcal{L}_{B}$, that the following quantity is conserved:
\BEA
    L &=& X\dot{Y} -Y\dot{X} 
    + {\sum}_{k=1}^N m_k\left(x_k \dot{y_k}-y_k\dot{x_k}\right) \nonumber\\
    &+& \frac{b}{2}\left(X^2+Y^2\right).
    \label{momentum}
\EEA
$L$ is a sum of the particle's angular momentum $X\dot{Y} -Y\dot{X}$ along $z$ direction, angular momenta of all bath oscillators and a contribution $\frac{b}{2}\left(X^2+Y^2\right)$ from the charged particle related to the magnetic field. Solving Euler-Lagrange equations of motion generated by $\mathcal{L}_{S}+\mathcal{L}_{B}$ for $\r_k$ 
we get: 
\BEA
    \r_k(t)=\r_k(0) \cos(\omega_k t) + \frac{\dot{\r}_{k}(0)}{\omega_k} \sin(\omega_k t)\nonumber\\
    +\frac{c_k}{m_k\omega_k} \int_0^t \d t' \sin(\omega_k (t-t')) \R(t').
    \label{osa}
\EEA
Plugging (\ref{osa}) into the Euler-Lagrange equations of motion for $\R$, we get the Langevin equation for the charged particle in magnetic field \cite{karmeshu} (see Appendix \ref{app-correlations}):
\BEA
\label{lang}
&&\ddot{\R}=b\ez\times\dot{\R}-\omega_0^2\R-\int_0^t\d u\zeta(t-u)\dot{\R}(u)+\bxi,\\
\label{kernel}
  &&  \zeta(t)={\sum}_{k=1}^N \frac{c_k^2}{m_k \omega_k^2} \cos(\omega_k t), \\ 
  \label{noise}
    && \bxi(t) = {\sum}_{k=1}^N[\, c_k\overline{\r}_k(0) \cos(\omega_k t)
    + \frac{c_k \v_{k}(0)}{\omega_k} \sin(\omega_k t)],\,\,~~\\
&&\overline{\r}_k(0)\equiv\r_k(0)-\R(0)\frac{ c_k}{m_k \omega_k^2}, \qquad \v_k=\dot{\r}_k. 
\label{ido}
\EEA
The cumulative force from the bath is decomposed into friction with a kernel $\zeta$ and noise $\bxi(t)$ that emerges due to the initially random state of the bath. Let the initial state of the particle and bath is given by the density
\BEA
{\cal P}(\r_k,\v_k,\R,\V) \propto e^{- H_B(\overline{\r}_k, {\v}_k)/T} {\cal P}(\R,\V),
~~ \V=\dot\R,~~~~
\label{ino}
\EEA
where $T$ is the temperature ($k_{B}=1$), and where the interaction-dressed bath energy $H_B$ reads from (\ref{lagr_bath}):
\BEA
\label{ham_bath}
H_B&=&-{\cal L}_B+{\sum}_{k=1}^N{m_k}\dot{\r}_k^2
\\
&=&\sum_{k=1}^N \left[\frac{m_k}{2}\dot{\r}_k^2 +
\frac{m_{k}\omega_{k}^{2}}{2} \left(\r_{k}-\frac{c_k\R}{m_k\omega_k^2}\right)^{2}\right].
\EEA
Eq.~(\ref{ino}) shows that bath's initial density is Gibbsian. Particle's initial density ${\cal P}(\R,\V)$ is arbitrary. The initial state (\ref{ino}) is not independent over the particle and bath, since $e^{- H_B(\overline{\r}_k, {\v}_k)/T}$ contains the shifted coordinate $\overline{\r}_k$; cf.~(\ref{ido}). Eq.~(\ref{ino}) refers to the time-scale separation, where the bath is prepared in equilibrium under fixed coordinate and momentum of the particle. This is realistic for Brownian motion.

Eqs.~(\ref{noise}, \ref{ino}) imply the fluctuation-dissipation relation
\begin{align}
    \la \xi_\alpha(t)\xi_\beta(t')\ra = T \delta_{\alpha\beta} \zeta(t-t') \qquad \alpha,\beta=x,y,z,
    \label{fdr}
\end{align}
where $\la...\ra$ is the average over initial state (\ref{ino}). 

The thermodynamic limit $N\to\infty$ for the bath is taken together with the dense frequency limit
$ \delta\omega\to 0$ ($n=1,...,N$) under condition $N\delta\omega\gg 1$ (see Appendix \ref{app-langevin} for details). Simultaneously, we take a weak coupling $c_n\to 0$ to each oscillator 
\cite{magal,petr,zwanzig,leggett,tuckerman}. The specific choice
\BEA
\label{oli}
&& \omega_n=\delta\omega\,n,~  c_n = \sqrt{{2\gamma \omega_n^2m_n \delta\omega}/{\pi}},\\
&& \zeta(t) =2 \gamma\,\delta(t),
\label{oligo}
\EEA
reproduces in the Langevin equation (\ref{lang}, \ref{kernel}, \ref{noise}) the Ohmic friction with magnitude $\gamma$ and the white noise \cite{leggett,petr,tuckerman,magal,zwanzig}.

\section{The \BvL theorem}
\label{sec-bvl}

Fluctuation-dissipation relation (\ref{fdr}) ensures from (\ref{lang})
(together with the thermodynamic limit for the bath) that the particle's state relaxes to the Gibbsian density \cite{magal,petr,zwanzig,leggett,tuckerman}:
\BEA
{\cal P}(\R,\V; t\to\infty)
\propto e^{-\frac{1}{2} ({\V}^2 +{\omega_0^2}\R^2)/T },
\label{bohr}
\EEA
that does not (and cannot) contain the magnetic field $b$. Obviously, the average angular momentum $\la XV_y-YV_x\ra$ calculated via (\ref{bohr}) is zero. Eq.~(\ref{bohr}) implies the \BvL theorem for the Brownian motion \cite{vleck,landau,kaplan,seifert}. The absence of the magnetic field $\bm{B}$ from the equilibrium density holds for any confining potential \cite{vleck,seifert}. This theorem is an equilibrium result and it does not hold for non-equilibrium steady states \cite{kumar1,kumar2,hidalgo,active2,active3,ashot}. In particular, the theorem may be broken by a weak white noise, which is sufficient for generating a sizable diamagnetic angular momentum for the Brownian particle \cite{ashot}. 

Note as well that the magnetic field shows up in  correlation functions of the Brownian particle and in its relaxation times to the equilibrium state \cite{ashot}; see also Appendix \ref{app-correlations}. This however is not the long-time response we seek here, since correlation functions decay within the relaxation time.

\section{Angular momentum of bath modes}
\label{sec-angular}

Let us assume that the initial state of the Brownian particle in (\ref{ino}) holds
\begin{gather}
{\cal P}(\R,\V)={\cal P}_1(X){\cal P}_1(Y){\cal P}_1(Z){\cal P}_2(V_x){\cal P}_2(V_y){\cal P}_2(V_z),\nonumber\\ 
{\cal P}_1(-a)={\cal P}_1(a),\qquad {\cal P}_2(-a)={\cal P}_2(a).    
\label{copt2}
\end{gather}
Eqs.~(\ref{copt2}) allow a large class of initial non-equilibrium states for the particle. Note that the initial angular momentum of the particle nullifies: $\la XV_y-YV_x \ra=0$. The virtue of (\ref{copt2}, \ref{ino}) is that all averages hold specific symmetry features related to the invariance of (\ref{copt2}) with respect to $\pi/2$ rotations in the $(x,y)$ plane; see Appendix \ref{app-symmetries}.

For initial second moments in (\ref{copt2}) we denote
\BEA
\label{or1}
 \la X^2\ra=\la Y^2\ra = \sigma_X \; {T}/{\omega_0^2},~~
 \la V_x^2\ra =\la V_y^2\ra = \sigma_V \; T,~~~
\EEA
where $\sigma_X=\sigma_V=1$ in (\ref{or1}) refers to equilibrium second moments of the initial state; cf.~(\ref{bohr}).

The long-time average angular momentum $L(\omega)$ of a bath oscillator with frequency $\omega=\omega_k$ is calculated from (\ref{osa}, \ref{lang}, \ref{copt2}) using the Laplace transform (see Appendix \ref{app-angular-momentum}):
\BEA
\label{orbi}
&&    L(\omega)\equiv m_k\la x_k v_{yk}-y_kv_{xk}\ra_{t\to\infty}=\delta\omega\,\frac{4\gamma b T}{\pi}\times\\ 
&&\frac{(\omega^2-\omega_0^2)\left((1-\sigma_V) \omega^{2}+(1-\sigma_X) \omega_0^{2}\right)}
    {\left((\omega^{2}-\omega_0^{2}-b\omega)^2+\gamma^2\omega^2 
    \right)
    \left((\omega^{2}-\omega_0^{2}+b\omega)^2+\gamma^2\omega^2\right)}.\nonumber
\EEA
Eq.~(\ref{orbi}) shows that even though bath oscillators are not charged, they acquire a non-zero angular momentum: $L(\omega)\not=0$ for $t\to\infty$, i.e. for times much larger than the relaxation time of the particle. This means that the long-time state of the bath feels the magnetic field $b$. Eqs.~(\ref{ino}, \ref{copt2}) show that the initial mean angular momentum nullifies both for the bath oscillator and the Brownian particle. Hence the fact of $L(\omega)\not=0$ for $t\to\infty$ means that the final state of the bath is out of equilibrium. 

Now $L(\omega_k)\propto c_k^2\propto \delta\omega$ since
$L(\omega_k)$ is driven by the coupling of the corresponding mode with the Brownian particle; cf.~(\ref{osa}). In view of (\ref{orbi}), the observable collective mode momentum is well-defined and amounts to integral between two finite frequencies:
\BEA
\sum_{k=k_1}^{k_2}L(\omega_k)=
\sum_{\omega=\omega_1}^{\omega_2}L(\omega)=\int_{\omega_1}^{\omega_2} \frac{\d\omega}{\delta\omega}\,L(\omega).
\label{it}
\EEA
Eq.~(\ref{orbi}) [and (\ref{tush}) below] can describe a finite-bath situation \cite{gemmer,olshanii,shaw,faria,riera}. Here the bath is large (for Langevin's equation (\ref{lang}) to apply for certain times), but finite. 

\begin{figure}[h!]
\includegraphics[width=0.8\columnwidth]{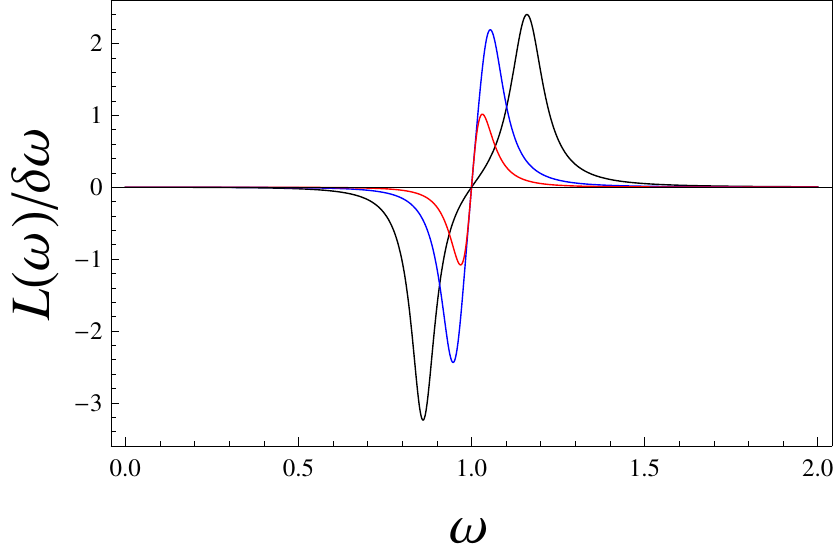}
\caption{\label{fig1} $L(\omega)/\delta\omega$ versus $\omega$ for 
$T=1$, $\omega_0=1$, $\gamma=0.1$, $\sigma_X=1$, $\sigma_V=0.1$; cf.~(\ref{orbi}, \ref{or1}). 
Curves refer to different magnetic fields (\ref{mag}) (from bottom to top): $b=0.3$ (black), $b=0.1$ (blue), $b=0.03$ (red).}
\end{figure}

We emphasize that $L(\omega)\not=0$ due to initially non-equilibrium state of the particle, since $L(\omega)=0$ whenever $\sigma_X=\sigma_V=1$ in (\ref{copt2}); cf.~(\ref{orbi}). Next, we confirm that (\ref{orbi}) is consistent with the average angular momentum conservation (\ref{momentum}) between $t=0$ and $t=\infty$:
\BEA
\int_{0}^{\infty}\d \omega\,L(\omega)=-\frac{Tb(1-\sigma_X)}{\omega_0^2}=b[\langle X^2 \rangle_0-\langle X^2 \rangle_\infty],~~~~
\EEA
where $\langle X^2 \rangle_0$ and $\langle X^2 \rangle_\infty$ are the initial and final values; cf.~(\ref{or1}). Even for $\sigma_X=1$ we can have $L(\omega)\not=0$ due to $\sigma_V\not=1$ in (\ref{orbi}). Hence, $L(\omega)\not=0$ need not be driven by the conservation law; see Fig.~\ref{fig1}. Now $L(\omega)$ changes its sign at $\omega=\omega_0$ and goes to zero as $\sim \omega^{-4}$ for $\omega\gg{\rm max}[\omega_0,\gamma,b]$; i.e. two sets of oscillators rotate in different directions, as Fig.~\ref{fig1} shows.

\section{Energy of bath modes}
\label{sec-energy}

Let us now see that the magnetic field also shows up in interaction-induced corrections to the energy of the bath mode. Eqs.~(\ref{ino}, \ref{ham_bath}) imply that the average energy of the bath mode with the frequency $\omega=\omega_k$ can be defined as 
\begin{align}
    E(\omega_k)=& \frac{m_k}{2} \la\dot{\r}_k^{2}\ra+\frac{m_k \omega_k^{2}}{2} \la{\r}_k^{2}\ra
    \nonumber \\
    &+\la \R^2\ra  \frac{c_k^{2}}{2m_k \omega_k^{2}} -c_k\la \R \cdot \r_k \ra,
    \label{colibri}
\end{align}
where the average is taken over the initial state (\ref{ino}). The full conserved energy of the particle+bath amounts to the sum of (\ref{colibri}) over all modes, $\sum_kE(\omega_k)$, plus the mean energy $\langle\frac{1}{2}\dot{\R}^2 +\frac{\omega_0^2}{2}\R^2\rangle$ of the particle; cf.~(\ref{lagr_sys}, \ref{ham_bath}). Eq.~(\ref{colibri}) in the limit $t\to\infty$ is worked out in Appendix \ref{app-energy}:
\BEA
\label{urum}
E(\omega)&=&3T 
-\delta\omega\left(2\varepsilon(\omega)+\varepsilon(\omega)|_{b=0} \right),
\\
\varepsilon(\omega)&=&\frac{T \gamma}{2\pi}\times 
\frac{ \omega^{2}(b^{2}+\gamma^{2})+\left(\omega^{2}-\omega_0^{2}\right)^{2}}
{(\omega^{2}-\omega_0^{2}-b\omega)^2+\gamma^2
\omega^2}\nonumber
\\
&&\qquad\times\frac{(1-\sigma_V) \omega^{2}+(1-\sigma_X) \omega_0^{2}}{(\omega^{2}-\omega_0^{2}+b\omega)^2+\gamma^2\omega^2}. \label{chi}
\EEA
where $\sigma_X$ and $\sigma_V$ are defined in (\ref{or1}). Now $3T$ in $E(\omega)$ is the thermal energy of the free mode; cf. the equipartition theorem from (\ref{ino}). The factor $\propto \delta\omega$ in (\ref{chi}) is due to interaction with the Brownian particle: $2\varepsilon(\omega)$ comes from $(x,y)$-components of (\ref{colibri}) that feel the magnetic field $b$ [cf.~(\ref{lang})], while $\varepsilon(\omega)|_{b=0}$ comes from the $z$-component that does not feel it. Similar to (\ref{orbi}), the interaction-induced factor in (\ref{chi}) nullifies for $\sigma_X=\sigma_V=1$; see (\ref{ino}, \ref{or1}). Thus, for long times the mode mean energy (\ref{chi}) (in contrast to particle's mean energy) depends on the magnetic field, though this dependence is weaker than for (\ref{orbi}). 
The interaction-driven factor in (\ref{urum}, \ref{chi}) that contains the magnetic field is a small correction to the leading term $2T$. This is different from (\ref{orbi}), where the average angular momentum without interaction is zero, and hence the magnetic field induced effect is the main one. 

\section{Linear momentum}
\label{sec-linear}

Given the above finding for the angular momentum, it is reasonable to ask how the linear momentum is transferred from the Brownian particle to the bath. Answering this question has intrinsic value and contrasts with the angular momentum's behavior.

Note from (\ref{lagr_bath}, \ref{lagr_sys}) that for a free Brownian particle (i.e. $\omega_0={\bf B}=0$) there is a conservation of linear momentum: 
\BEA
\frac{\d }{\d t}\left[{\sum}_{k=1}^N\,\,
\frac{c_k}{\omega_k^2}\, \dot\r_k(t)+\V(t)
\right]=0.
\label{gnu_main}
\EEA
Eq.~(\ref{gnu_main}) contains the full momentum $\V(t)$ of the Brownian particle (recall that its mass is taken $1$), which is added to $\frac{c_k}{\omega_k^2}\, \dot\r_k(t)$ for each mode. The latter is a part of the mode momentum $m_k\dot\r_k$. We emphasize that the full mode momentum $m_k\dot\r_k$ cannot appear in the conservation law, since the model lacks the full translation invariance. Indeed, (\ref{lagr_bath}) shows that each mode feels a harmonic potential. A relations similar to (\ref{gnu_main}) is formally mentioned in Ref.~\cite{hakim}.

Instead of (\ref{copt2}, \ref{or1}) consider initial conditions $\langle\R(0) \rangle=0$, $\langle{\V}(0) \rangle\not=0$ that hold together with (\ref{ino}). As follows from (\ref{lang}, \ref{oligo}), and is generally well-known, the velocity density of the particle thermalizes for long times [cf.~(\ref{bohr})]: 
\BEA
{\cal P}(\V; t\to\infty)\propto e^{-\frac{1}{2T} {\V}^2}.
\label{toros}
\EEA
But $\R$ is subject to an unbound Brownian motion and hence it is not thermalized. 

For $\gamma t\gg 1$ we find from (\ref{osa}, \ref{lang}, \ref{oligo}, \ref{ino}) for the average linear momentum of a bath mode with frequency $\omega=\omega_k$ (see Appendix \ref{app-momentum}):
\begin{align}
    \frac{c_k}{\omega_k^2}\langle\dot\r_k\rangle
    = \delta\omega\times \frac{2\gamma\langle \V(0)\rangle}{\pi}\,\,
\frac{\frac{\gamma}{\omega} \sin[\omega t]- \cos[\omega t]}{\gamma^2+\omega^2}.
\label{tush}
\end{align}
Eq.~(\ref{tush}) shows that the single mode (partial) linear momentum $    \frac{c_k}{\omega_k^2}\langle\dot\r_k\rangle$ is a time-dependent oscillating function. 
In contrast, relaxation occurs once we consider a collective quantity and take the bath's thermodynamic limit, i.e. using the integral instead of the sum:
\BEA
\sum_{\omega_1}^{\omega_2}\frac{c_k}{\omega_k^2}\langle\dot\r_k\rangle=\frac{2\gamma\langle \V(0)\rangle}{\pi}
\int_{\omega_1}^{\omega_2}\d\omega\,\frac{\frac{\gamma}{\omega} \sin[\omega t]- \cos[\omega t]}{\gamma^2+\omega^2},~~~~
\label{lorbi}
\EEA
where we used (\ref{tush}), and where naturally $\omega_2>\omega_1$.

The behavior of the integral in (\ref{lorbi}) for $t\to\infty$ essentially depends on whether $\omega_1>0$ or $\omega_1=0$, i.e. whether the zero frequency is included in the collective quantity or not. 
For $\omega_1>0$, (\ref{lorbi}) implies
\BEA
\sum_{\omega_1>0}^{\omega_2}\frac{c_k}{\omega_k^2}\langle\dot\r_k\rangle=O\left(\frac{1}{t}\right)\quad {\rm for}\quad
t\to\infty,
\label{lo}
\EEA
as becomes clear after changing the variable $\omega t\to\omega$ in (\ref{lorbi}). Hence the collective linear momentum for non-zero frequencies nullifies for large times. Note that these times are larger than $1/\gamma$. 

However, the total bath linear momentum is non-zero, as follows from (\ref{gnu_main}). Indeed, applying (\ref{gnu_main}) at times $t=0$ and $t\gg 1/\gamma$ together with conditions $\langle\dot\r_k \rangle=0$, $\langle{\V}(0) \rangle\not=0$ and $\langle{\V}(t) \rangle \approxeq 0$ we find for the total final linear momentum:
\BEA
\sum_{k=1}^N\frac{c_k}{\omega_k^2}\, \dot\r_k(t)
&=&\frac{2\gamma\langle \V(0)\rangle}{\pi}
\int_{0}^{\infty}\d\omega\,\frac{\frac{\gamma}{\omega} \sin[\omega t]- \cos[\omega t]}{\gamma^2+\omega^2}\nonumber\\
&=&\langle \V(0)\rangle. 
\label{gn}
\EEA
Note that if $\omega_1$ is extended to zero in (\ref{lorbi}), we get back from the $t\to\infty$-limit of (\ref{lorbi}) the same total momentum $\langle \V(0)\rangle$. This means that the total momentum $\langle \V(0)\rangle$ is (for $t\to\infty$) transferred to the zero-frequency mode $\omega=0$, which corresponds to the homogeneous shift. Thus, the linear momentum dissipates away from both the particle velocity and collective bath observables $\sum_{\omega_1>0}^{\omega_2}\frac{c_k}{\omega_k^2}\langle\dot\r_k\rangle$. However, the velocity $\langle\dot\r_{\omega\to 0}\rangle$ of the zero-frequency mode is not observable, because it is infinitisemaly small due to a ${\rm lim}_{\omega\to 0}\frac{c_{\omega}}{\omega^2}=\infty$; see (\ref{oli}). Hence the total linear momentum does not result into any visible motion for long times. 

The comparison between the linear momentum and the angular momentum can be carried out in several different set-ups. We choose the most reasonable set-up for showing that the linear momentum is different from the angular momentum. Within this set-up the behavior of the angular momentum for the Brownian particle subject to the external potential with frequency $\omega_0\not=0$ and magnetic field ${\bf B}\not=0$ is compared with the linear momentum under $\omega_0=0$ and ${\bf B}=0$; see (\ref{gnu_main}--\ref{gn}). In the former case the overall angular momentum is conserved [cf.~(\ref{momentum})], while the final state of the Brownian particle thermalizes; see (\ref{bohr}). In the latter case, the partial linear momentum is conserved, and there is a partial long-time thermalization; see (\ref{gnu_main}) and (\ref{toros}) below. Other possible set-up for comparing the linear and angular momentum are discussed in Appendix \ref{app-momentum}; see Table \ref{tabu} there.

There are two major differences between the angular momentum and linear momentum. First, the single mode angular momentum (\ref{orbi}) converges to a well-defined, time-independent value, in contrast to the single mode linear momentum (\ref{tush}) that is an oscillating function. Consequently, the angular momentum is sustained in the long-time limit for finite bath observables in the thermodynamic limit, while the linear momentum dissipates away; cf.~(\ref{lo}) with (\ref{it}). Second, (\ref{tush}, \ref{lorbi}) trivialize for zero initial momentum $\langle \V(0)\rangle=0$, which again contrasts (\ref{orbi}) that does not need an initial angular momentum for the particle. 

\section{Summary}

We restored the influence of the static (homogeneous) magnetic field on the long-time limit of a classical Brownian motion. For long times the influence of the magnetic field is found not in the particle (as correctly claimed by \BvL theorem), but in the orbital momentum of (uncharged) bath modes. The effect is not enforced by the conservation law of the angular momentum, i.e. there are situations where its absence is fully consistent with the conservation. This behavior is specific for the angular momentum, as compared to the two other additive integrals of motion: the linear momentum is dissipated away by (and from) bath modes, while the bath energy can feel the magnetic field, but only as a small correction. 

At this point, the major open problem is whether these effects are specific for the Caldeira-Leggett model of the bath|which we emphasize does apply for describing liquids \cite{stratt,tuckerman}|or it will generalize to more realistic bath models. This question is currently under investigation, but we can mention two preliminary hypotheses. The non-spontaneous aspect of the bath angular momentum is based on the conservation law (\ref{momentum}). Hence it is likely to persist for more general bath models. The spontaneous aspect could be a specific point of linear bath models, including the Caldeira-Leggett model. 

In this context, we plan to study bath models that are especially relevant for cellular ions. Here the bath is described as a fluctuating hydrodynamic system \cite{lp}. It is known that such a bath also reproduces the Langevin equation for the Brownian particle immersed into it \cite{mazur}. Hence we anticipate that the effect of transferring the influence of the magnetic field to the bath will show up also here, and will be reflected in the angular momentum of the fluid. The major difference of this class of models compared to the Caldeira-Leggett model is that bath modes are subject to viscosity, i.e. there is an additional source of irreversibility. 

\acknowledgments
This work was supported by SCS of Armenia grants No. 21AG-1C038, No. 22AA-1C028 and No. 20TTAT-QTa003. We thank David Petrosyan for important remarks.

\bibliography{pre}

\begin{thebibliography}{39}%
\makeatletter
\providecommand \@ifxundefined [1]{%
 \@ifx{#1\undefined}
}%
\providecommand \@ifnum [1]{%
 \ifnum #1\expandafter \@firstoftwo
 \else \expandafter \@secondoftwo
 \fi
}%
\providecommand \@ifx [1]{%
 \ifx #1\expandafter \@firstoftwo
 \else \expandafter \@secondoftwo
 \fi
}%
\providecommand \natexlab [1]{#1}%
\providecommand \enquote  [1]{``#1''}%
\providecommand \bibnamefont  [1]{#1}%
\providecommand \bibfnamefont [1]{#1}%
\providecommand \citenamefont [1]{#1}%
\providecommand \href@noop [0]{\@secondoftwo}%
\providecommand \href [0]{\begingroup \@sanitize@url \@href}%
\providecommand \@href[1]{\@@startlink{#1}\@@href}%
\providecommand \@@href[1]{\endgroup#1\@@endlink}%
\providecommand \@sanitize@url [0]{\catcode `\\12\catcode `\$12\catcode
  `\&12\catcode `\#12\catcode `\^12\catcode `\_12\catcode `\%12\relax}%
\providecommand \@@startlink[1]{}%
\providecommand \@@endlink[0]{}%
\providecommand \url  [0]{\begingroup\@sanitize@url \@url }%
\providecommand \@url [1]{\endgroup\@href {#1}{\urlprefix }}%
\providecommand \urlprefix  [0]{URL }%
\providecommand \Eprint [0]{\href }%
\providecommand \doibase [0]{https://doi.org/}%
\providecommand \selectlanguage [0]{\@gobble}%
\providecommand \bibinfo  [0]{\@secondoftwo}%
\providecommand \bibfield  [0]{\@secondoftwo}%
\providecommand \translation [1]{[#1]}%
\providecommand \BibitemOpen [0]{}%
\providecommand \bibitemStop [0]{}%
\providecommand \bibitemNoStop [0]{.\EOS\space}%
\providecommand \EOS [0]{\spacefactor3000\relax}%
\providecommand \BibitemShut  [1]{\csname bibitem#1\endcsname}%
\let\auto@bib@innerbib\@empty
\bibitem [{\citenamefont {Van~Vleck}(1965)}]{vleck}%
  \BibitemOpen
  \bibfield  {author} {\bibinfo {author} {\bibfnamefont {J.~H.}\ \bibnamefont
  {Van~Vleck}},\ }\href@noop {} {\emph {\bibinfo {title} {The theory of
  electric and magnetic susceptibilities}}}\ (\bibinfo  {publisher} {Oxford
  University Press},\ \bibinfo {year} {1965})\BibitemShut {NoStop}%
\bibitem [{\citenamefont {Kaplan}\ and\ \citenamefont
  {Mahanti}(2009)}]{kaplan}%
  \BibitemOpen
  \bibfield  {author} {\bibinfo {author} {\bibfnamefont {T.}~\bibnamefont
  {Kaplan}}\ and\ \bibinfo {author} {\bibfnamefont {S.}~\bibnamefont
  {Mahanti}},\ }\bibfield  {title} {\bibinfo {title} {On the bohr-van leeuwen
  theorem, the non-existence of classical magnetism in thermal equilibrium},\
  }\href@noop {} {\bibfield  {journal} {\bibinfo  {journal} {EPL (Europhysics
  Letters)}\ }\textbf {\bibinfo {volume} {87}},\ \bibinfo {pages} {17002}
  (\bibinfo {year} {2009})}\BibitemShut {NoStop}%
\bibitem [{\citenamefont {Pradhan}\ and\ \citenamefont
  {Seifert}(2010)}]{seifert}%
  \BibitemOpen
  \bibfield  {author} {\bibinfo {author} {\bibfnamefont {P.}~\bibnamefont
  {Pradhan}}\ and\ \bibinfo {author} {\bibfnamefont {U.}~\bibnamefont
  {Seifert}},\ }\bibfield  {title} {\bibinfo {title} {Nonexistence of classical
  diamagnetism and nonequilibrium fluctuation theorems for charged particles on
  a curved surface},\ }\href@noop {} {\bibfield  {journal} {\bibinfo  {journal}
  {EPL (Europhysics Letters)}\ }\textbf {\bibinfo {volume} {89}},\ \bibinfo
  {pages} {37001} (\bibinfo {year} {2010})}\BibitemShut {NoStop}%
\bibitem [{\citenamefont {Landau}\ and\ \citenamefont
  {Lifshitz}(1978)}]{landau}%
  \BibitemOpen
  \bibfield  {author} {\bibinfo {author} {\bibfnamefont {L.}~\bibnamefont
  {Landau}}\ and\ \bibinfo {author} {\bibfnamefont {E.}~\bibnamefont
  {Lifshitz}},\ }\href@noop {} {\emph {\bibinfo {title} {Statistical physics,
  I. Course of Theoretical physics: Volume 5}}}\ (\bibinfo  {publisher}
  {Oxford: Pergamon},\ \bibinfo {year} {1978})\BibitemShut {NoStop}%
\bibitem [{\citenamefont {Weibel}(1959)}]{weibel}%
  \BibitemOpen
  \bibfield  {author} {\bibinfo {author} {\bibfnamefont {E.~S.}\ \bibnamefont
  {Weibel}},\ }\bibfield  {title} {\bibinfo {title} {On the confinement of a
  plasma by magnetostatic fields},\ }\href@noop {} {\bibfield  {journal}
  {\bibinfo  {journal} {The Physics of Fluids}\ }\textbf {\bibinfo {volume}
  {2}},\ \bibinfo {pages} {52} (\bibinfo {year} {1959})}\BibitemShut {NoStop}%
\bibitem [{\citenamefont {Ivlev}\ \emph {et~al.}(2012)\citenamefont {Ivlev},
  \citenamefont {Morfill}, \citenamefont {Lowen},\ and\ \citenamefont
  {Royall}}]{book_plasma_colloid}%
  \BibitemOpen
  \bibfield  {author} {\bibinfo {author} {\bibfnamefont {A.}~\bibnamefont
  {Ivlev}}, \bibinfo {author} {\bibfnamefont {G.}~\bibnamefont {Morfill}},
  \bibinfo {author} {\bibfnamefont {H.}~\bibnamefont {Lowen}},\ and\ \bibinfo
  {author} {\bibfnamefont {C.~P.}\ \bibnamefont {Royall}},\ }\href@noop {}
  {\emph {\bibinfo {title} {Complex plasmas and colloidal dispersions:
  particle-resolved studies of classical liquids and solids}}},\ Vol.~\bibinfo
  {volume} {5}\ (\bibinfo  {publisher} {World Scientific Publishing Company},\
  \bibinfo {year} {2012})\BibitemShut {NoStop}%
\bibitem [{\citenamefont {K{\"a}hlert}\ \emph {et~al.}(2012)\citenamefont
  {K{\"a}hlert}, \citenamefont {Carstensen}, \citenamefont {Bonitz},
  \citenamefont {L{\"o}wen}, \citenamefont {Greiner},\ and\ \citenamefont
  {Piel}}]{plasma_rotation}%
  \BibitemOpen
  \bibfield  {author} {\bibinfo {author} {\bibfnamefont {H.}~\bibnamefont
  {K{\"a}hlert}}, \bibinfo {author} {\bibfnamefont {J.}~\bibnamefont
  {Carstensen}}, \bibinfo {author} {\bibfnamefont {M.}~\bibnamefont {Bonitz}},
  \bibinfo {author} {\bibfnamefont {H.}~\bibnamefont {L{\"o}wen}}, \bibinfo
  {author} {\bibfnamefont {F.}~\bibnamefont {Greiner}},\ and\ \bibinfo {author}
  {\bibfnamefont {A.}~\bibnamefont {Piel}},\ }\bibfield  {title} {\bibinfo
  {title} {Magnetizing a complex plasma without a magnetic field},\ }\href@noop
  {} {\bibfield  {journal} {\bibinfo  {journal} {Physical review letters}\
  }\textbf {\bibinfo {volume} {109}},\ \bibinfo {pages} {155003} (\bibinfo
  {year} {2012})}\BibitemShut {NoStop}%
\bibitem [{\citenamefont {L{\"o}wen}(2013)}]{lowen}%
  \BibitemOpen
  \bibfield  {author} {\bibinfo {author} {\bibfnamefont {H.}~\bibnamefont
  {L{\"o}wen}},\ }\bibfield  {title} {\bibinfo {title} {Introduction to
  colloidal dispersions in external fields},\ }\href@noop {} {\bibfield
  {journal} {\bibinfo  {journal} {The European Physical Journal Special
  Topics}\ }\textbf {\bibinfo {volume} {222}},\ \bibinfo {pages} {2727}
  (\bibinfo {year} {2013})}\BibitemShut {NoStop}%
\bibitem [{\citenamefont {Reeves}\ \emph {et~al.}(2018)\citenamefont {Reeves},
  \citenamefont {Aronson},\ and\ \citenamefont {Vlahovska}}]{aronson}%
  \BibitemOpen
  \bibfield  {author} {\bibinfo {author} {\bibfnamefont {C.}~\bibnamefont
  {Reeves}}, \bibinfo {author} {\bibfnamefont {I.}~\bibnamefont {Aronson}},\
  and\ \bibinfo {author} {\bibfnamefont {P.}~\bibnamefont {Vlahovska}},\
  }\bibfield  {title} {\bibinfo {title} {Active suspension of self-rotating
  particles},\ }\href@noop {} {\bibfield  {journal} {\bibinfo  {journal}
  {Bulletin of the American Physical Society}\ }\textbf {\bibinfo {volume}
  {63}} (\bibinfo {year} {2018})}\BibitemShut {NoStop}%
\bibitem [{\citenamefont {Zhang}\ \emph {et~al.}(2017)\citenamefont {Zhang},
  \citenamefont {Yarema},\ and\ \citenamefont {Xu}}]{zarema}%
  \BibitemOpen
  \bibfield  {author} {\bibinfo {author} {\bibfnamefont {X.}~\bibnamefont
  {Zhang}}, \bibinfo {author} {\bibfnamefont {K.}~\bibnamefont {Yarema}},\ and\
  \bibinfo {author} {\bibfnamefont {A.}~\bibnamefont {Xu}},\ }\href@noop {}
  {\emph {\bibinfo {title} {Biological effects of static magnetic fields}}}\
  (\bibinfo  {publisher} {Springer},\ \bibinfo {year} {2017})\BibitemShut
  {NoStop}%
\bibitem [{\citenamefont {Albuquerque}\ \emph {et~al.}(2016)\citenamefont
  {Albuquerque}, \citenamefont {Costa}, \citenamefont {e~Fernandes},\ and\
  \citenamefont {Porto}}]{albu}%
  \BibitemOpen
  \bibfield  {author} {\bibinfo {author} {\bibfnamefont {W.~W.~C.}\
  \bibnamefont {Albuquerque}}, \bibinfo {author} {\bibfnamefont {R.~M. P.~B.}\
  \bibnamefont {Costa}}, \bibinfo {author} {\bibfnamefont {T.~d.~S.}\
  \bibnamefont {e~Fernandes}},\ and\ \bibinfo {author} {\bibfnamefont
  {A.~L.~F.}\ \bibnamefont {Porto}},\ }\bibfield  {title} {\bibinfo {title}
  {Evidences of the static magnetic field influence on cellular systems},\
  }\href@noop {} {\bibfield  {journal} {\bibinfo  {journal} {Progress in
  Biophysics and Molecular Biology}\ }\textbf {\bibinfo {volume} {121}},\
  \bibinfo {pages} {16} (\bibinfo {year} {2016})}\BibitemShut {NoStop}%
\bibitem [{\citenamefont {Adair}(2000)}]{adair}%
  \BibitemOpen
  \bibfield  {author} {\bibinfo {author} {\bibfnamefont {R.~K.}\ \bibnamefont
  {Adair}},\ }\bibfield  {title} {\bibinfo {title} {Static and low-frequency
  magnetic field effects: health risks and therapies},\ }\href@noop {}
  {\bibfield  {journal} {\bibinfo  {journal} {Reports on progress in Physics}\
  }\textbf {\bibinfo {volume} {63}},\ \bibinfo {pages} {415} (\bibinfo {year}
  {2000})}\BibitemShut {NoStop}%
\bibitem [{\citenamefont {Glaser}(2012)}]{glaser}%
  \BibitemOpen
  \bibfield  {author} {\bibinfo {author} {\bibfnamefont {R.}~\bibnamefont
  {Glaser}},\ }\href@noop {} {\emph {\bibinfo {title} {Biophysics: an
  introduction}}}\ (\bibinfo  {publisher} {Springer Science \& Business
  Media},\ \bibinfo {year} {2012})\BibitemShut {NoStop}%
\bibitem [{\citenamefont {Bialek}(2012)}]{bialek}%
  \BibitemOpen
  \bibfield  {author} {\bibinfo {author} {\bibfnamefont {W.}~\bibnamefont
  {Bialek}},\ }\href@noop {} {\emph {\bibinfo {title} {Biophysics: searching
  for principles}}}\ (\bibinfo  {publisher} {Princeton University Press},\
  \bibinfo {year} {2012})\BibitemShut {NoStop}%
\bibitem [{\citenamefont {Tiersch}\ and\ \citenamefont
  {Briegel}(2012)}]{briegel}%
  \BibitemOpen
  \bibfield  {author} {\bibinfo {author} {\bibfnamefont {M.}~\bibnamefont
  {Tiersch}}\ and\ \bibinfo {author} {\bibfnamefont {H.~J.}\ \bibnamefont
  {Briegel}},\ }\bibfield  {title} {\bibinfo {title} {Decoherence in the
  chemical compass: the role of decoherence for avian magnetoreception},\
  }\href@noop {} {\bibfield  {journal} {\bibinfo  {journal} {Philosophical
  Transactions of the Royal Society A: Mathematical, Physical and Engineering
  Sciences}\ }\textbf {\bibinfo {volume} {370}},\ \bibinfo {pages} {4517}
  (\bibinfo {year} {2012})}\BibitemShut {NoStop}%
\bibitem [{\citenamefont {Cai}\ and\ \citenamefont {Plenio}(2013)}]{cai}%
  \BibitemOpen
  \bibfield  {author} {\bibinfo {author} {\bibfnamefont {J.}~\bibnamefont
  {Cai}}\ and\ \bibinfo {author} {\bibfnamefont {M.~B.}\ \bibnamefont
  {Plenio}},\ }\bibfield  {title} {\bibinfo {title} {Chemical compass model for
  avian magnetoreception as a quantum coherent device},\ }\href@noop {}
  {\bibfield  {journal} {\bibinfo  {journal} {Physical review letters}\
  }\textbf {\bibinfo {volume} {111}},\ \bibinfo {pages} {230503} (\bibinfo
  {year} {2013})}\BibitemShut {NoStop}%
\bibitem [{\citenamefont {Kominis}(2009)}]{kominis}%
  \BibitemOpen
  \bibfield  {author} {\bibinfo {author} {\bibfnamefont {I.~K.}\ \bibnamefont
  {Kominis}},\ }\bibfield  {title} {\bibinfo {title} {Quantum zeno effect
  explains magnetic-sensitive radical-ion-pair reactions},\ }\href@noop {}
  {\bibfield  {journal} {\bibinfo  {journal} {Physical Review E}\ }\textbf
  {\bibinfo {volume} {80}},\ \bibinfo {pages} {056115} (\bibinfo {year}
  {2009})}\BibitemShut {NoStop}%
\bibitem [{\citenamefont {Magalinskii}(1959)}]{magal}%
  \BibitemOpen
  \bibfield  {author} {\bibinfo {author} {\bibfnamefont {V.}~\bibnamefont
  {Magalinskii}},\ }\bibfield  {title} {\bibinfo {title} {Dynamical model in
  the theory of the brownian motion},\ }\href@noop {} {\bibfield  {journal}
  {\bibinfo  {journal} {Soviet Physics JETP}\ }\textbf {\bibinfo {volume}
  {9}},\ \bibinfo {pages} {1381} (\bibinfo {year} {1959})}\BibitemShut
  {NoStop}%
\bibitem [{\citenamefont {Zwanzig}(1973)}]{zwanzig}%
  \BibitemOpen
  \bibfield  {author} {\bibinfo {author} {\bibfnamefont {R.}~\bibnamefont
  {Zwanzig}},\ }\bibfield  {title} {\bibinfo {title} {Nonlinear generalized
  langevin equations},\ }\href@noop {} {\bibfield  {journal} {\bibinfo
  {journal} {Journal of Statistical Physics}\ }\textbf {\bibinfo {volume}
  {9}},\ \bibinfo {pages} {215} (\bibinfo {year} {1973})}\BibitemShut {NoStop}%
\bibitem [{\citenamefont {Caldeira}\ and\ \citenamefont
  {Leggett}(1983)}]{leggett}%
  \BibitemOpen
  \bibfield  {author} {\bibinfo {author} {\bibfnamefont {A.~O.}\ \bibnamefont
  {Caldeira}}\ and\ \bibinfo {author} {\bibfnamefont {A.~J.}\ \bibnamefont
  {Leggett}},\ }\bibfield  {title} {\bibinfo {title} {Quantum tunnelling in a
  dissipative system},\ }\href@noop {} {\bibfield  {journal} {\bibinfo
  {journal} {Annals of physics}\ }\textbf {\bibinfo {volume} {149}},\ \bibinfo
  {pages} {374} (\bibinfo {year} {1983})}\BibitemShut {NoStop}%
\bibitem [{\citenamefont {Breuer}\ \emph {et~al.}(2002)\citenamefont {Breuer},
  \citenamefont {Petruccione} \emph {et~al.}}]{petr}%
  \BibitemOpen
  \bibfield  {author} {\bibinfo {author} {\bibfnamefont {H.-P.}\ \bibnamefont
  {Breuer}}, \bibinfo {author} {\bibfnamefont {F.}~\bibnamefont {Petruccione}},
  \emph {et~al.},\ }\href@noop {} {\emph {\bibinfo {title} {The theory of open
  quantum systems}}}\ (\bibinfo  {publisher} {Oxford University Press on
  Demand},\ \bibinfo {year} {2002})\BibitemShut {NoStop}%
\bibitem [{\citenamefont {Tuckerman}(2010)}]{tuckerman}%
  \BibitemOpen
  \bibfield  {author} {\bibinfo {author} {\bibfnamefont {M.}~\bibnamefont
  {Tuckerman}},\ }\href@noop {} {\emph {\bibinfo {title} {Statistical
  mechanics: theory and molecular simulation}}}\ (\bibinfo  {publisher} {Oxford
  university press},\ \bibinfo {year} {2010})\BibitemShut {NoStop}%
\bibitem [{\citenamefont {Goodyear}\ \emph {et~al.}(1996)\citenamefont
  {Goodyear}, \citenamefont {Larsen},\ and\ \citenamefont
  {Stratt}}]{molecular}%
  \BibitemOpen
  \bibfield  {author} {\bibinfo {author} {\bibfnamefont {G.}~\bibnamefont
  {Goodyear}}, \bibinfo {author} {\bibfnamefont {R.~E.}\ \bibnamefont
  {Larsen}},\ and\ \bibinfo {author} {\bibfnamefont {R.~M.}\ \bibnamefont
  {Stratt}},\ }\bibfield  {title} {\bibinfo {title} {Molecular origin of
  friction in liquids},\ }\href@noop {} {\bibfield  {journal} {\bibinfo
  {journal} {Physical review letters}\ }\textbf {\bibinfo {volume} {76}},\
  \bibinfo {pages} {243} (\bibinfo {year} {1996})}\BibitemShut {NoStop}%
\bibitem [{\citenamefont {Stratt}(1995)}]{stratt}%
  \BibitemOpen
  \bibfield  {author} {\bibinfo {author} {\bibfnamefont {R.~M.}\ \bibnamefont
  {Stratt}},\ }\bibfield  {title} {\bibinfo {title} {The instantaneous normal
  modes of liquids},\ }\href@noop {} {\bibfield  {journal} {\bibinfo  {journal}
  {Accounts of Chemical Research}\ }\textbf {\bibinfo {volume} {28}},\ \bibinfo
  {pages} {201} (\bibinfo {year} {1995})}\BibitemShut {NoStop}%
\bibitem [{\citenamefont {Karmeshu}(1974)}]{karmeshu}%
  \BibitemOpen
  \bibfield  {author} {\bibinfo {author} {\bibnamefont {Karmeshu}},\ }\bibfield
   {title} {\bibinfo {title} {Brownian motion of charged particles in a
  magnetic field},\ }\href@noop {} {\bibfield  {journal} {\bibinfo  {journal}
  {The Physics of Fluids}\ }\textbf {\bibinfo {volume} {17}},\ \bibinfo {pages}
  {1828} (\bibinfo {year} {1974})}\BibitemShut {NoStop}%
\bibitem [{\citenamefont {Jayannavar}\ and\ \citenamefont
  {Kumar}(1981)}]{kumar1}%
  \BibitemOpen
  \bibfield  {author} {\bibinfo {author} {\bibfnamefont {A.}~\bibnamefont
  {Jayannavar}}\ and\ \bibinfo {author} {\bibfnamefont {N.}~\bibnamefont
  {Kumar}},\ }\bibfield  {title} {\bibinfo {title} {Orbital diamagnetism of a
  charged brownian particle undergoing a birth-death process},\ }\href@noop {}
  {\bibfield  {journal} {\bibinfo  {journal} {Journal of Physics A:
  Mathematical and General}\ }\textbf {\bibinfo {volume} {14}},\ \bibinfo
  {pages} {1399} (\bibinfo {year} {1981})}\BibitemShut {NoStop}%
\bibitem [{\citenamefont {Kumar}(2012)}]{kumar2}%
  \BibitemOpen
  \bibfield  {author} {\bibinfo {author} {\bibfnamefont {N.}~\bibnamefont
  {Kumar}},\ }\bibfield  {title} {\bibinfo {title} {Classical orbital magnetic
  moment in a dissipative stochastic system},\ }\href@noop {} {\bibfield
  {journal} {\bibinfo  {journal} {Physical Review E}\ }\textbf {\bibinfo
  {volume} {85}},\ \bibinfo {pages} {011114} (\bibinfo {year}
  {2012})}\BibitemShut {NoStop}%
\bibitem [{\citenamefont {Hidalgo-Gonzalez}\ \emph {et~al.}(2016)\citenamefont
  {Hidalgo-Gonzalez}, \citenamefont {Jim{\'e}nez-Aquino},\ and\ \citenamefont
  {Romero-Bastida}}]{hidalgo}%
  \BibitemOpen
  \bibfield  {author} {\bibinfo {author} {\bibfnamefont {J.}~\bibnamefont
  {Hidalgo-Gonzalez}}, \bibinfo {author} {\bibfnamefont {J.}~\bibnamefont
  {Jim{\'e}nez-Aquino}},\ and\ \bibinfo {author} {\bibfnamefont
  {M.}~\bibnamefont {Romero-Bastida}},\ }\bibfield  {title} {\bibinfo {title}
  {Non-markovian brownian motion in a magnetic field and time-dependent force
  fields},\ }\href@noop {} {\bibfield  {journal} {\bibinfo  {journal} {Physica
  A: Statistical Mechanics and its Applications}\ }\textbf {\bibinfo {volume}
  {462}},\ \bibinfo {pages} {1128} (\bibinfo {year} {2016})}\BibitemShut
  {NoStop}%
\bibitem [{\citenamefont {Abdoli}\ and\ \citenamefont
  {Sharma}(2021)}]{active2}%
  \BibitemOpen
  \bibfield  {author} {\bibinfo {author} {\bibfnamefont {I.}~\bibnamefont
  {Abdoli}}\ and\ \bibinfo {author} {\bibfnamefont {A.}~\bibnamefont
  {Sharma}},\ }\bibfield  {title} {\bibinfo {title} {Stochastic resetting of
  active brownian particles with lorentz force},\ }\href@noop {} {\bibfield
  {journal} {\bibinfo  {journal} {Soft Matter}\ }\textbf {\bibinfo {volume}
  {17}},\ \bibinfo {pages} {1307} (\bibinfo {year} {2021})}\BibitemShut
  {NoStop}%
\bibitem [{\citenamefont {Abdoli}\ \emph {et~al.}(2020)\citenamefont {Abdoli},
  \citenamefont {Vuijk}, \citenamefont {Wittmann}, \citenamefont {Sommer},
  \citenamefont {Brader},\ and\ \citenamefont {Sharma}}]{active3}%
  \BibitemOpen
  \bibfield  {author} {\bibinfo {author} {\bibfnamefont {I.}~\bibnamefont
  {Abdoli}}, \bibinfo {author} {\bibfnamefont {H.}~\bibnamefont {Vuijk}},
  \bibinfo {author} {\bibfnamefont {R.}~\bibnamefont {Wittmann}}, \bibinfo
  {author} {\bibfnamefont {J.}~\bibnamefont {Sommer}}, \bibinfo {author}
  {\bibfnamefont {J.}~\bibnamefont {Brader}},\ and\ \bibinfo {author}
  {\bibfnamefont {A.}~\bibnamefont {Sharma}},\ }\bibfield  {title} {\bibinfo
  {title} {Stationary state in brownian systems with lorentz force},\
  }\href@noop {} {\bibfield  {journal} {\bibinfo  {journal} {Physical Review
  Research}\ }\textbf {\bibinfo {volume} {2}},\ \bibinfo {pages} {023381}
  (\bibinfo {year} {2020})}\BibitemShut {NoStop}%
\bibitem [{\citenamefont {Matevosyan}\ and\ \citenamefont
  {Allahverdyan}(2021)}]{ashot}%
  \BibitemOpen
  \bibfield  {author} {\bibinfo {author} {\bibfnamefont {A.}~\bibnamefont
  {Matevosyan}}\ and\ \bibinfo {author} {\bibfnamefont {A.~E.}\ \bibnamefont
  {Allahverdyan}},\ }\bibfield  {title} {\bibinfo {title} {Nonequilibrium,
  weak-field-induced cyclotron motion: A mechanism for magnetobiology},\
  }\href@noop {} {\bibfield  {journal} {\bibinfo  {journal} {Physical Review
  E}\ }\textbf {\bibinfo {volume} {104}},\ \bibinfo {pages} {064407} (\bibinfo
  {year} {2021})}\BibitemShut {NoStop}%
\bibitem [{\citenamefont {Gemmer}\ and\ \citenamefont {Michel}(2005)}]{gemmer}%
  \BibitemOpen
  \bibfield  {author} {\bibinfo {author} {\bibfnamefont {J.}~\bibnamefont
  {Gemmer}}\ and\ \bibinfo {author} {\bibfnamefont {M.}~\bibnamefont
  {Michel}},\ }\bibfield  {title} {\bibinfo {title} {Thermalization of quantum
  systems by finite baths},\ }\href@noop {} {\bibfield  {journal} {\bibinfo
  {journal} {EPL (Europhysics Letters)}\ }\textbf {\bibinfo {volume} {73}},\
  \bibinfo {pages} {1} (\bibinfo {year} {2005})}\BibitemShut {NoStop}%
\bibitem [{\citenamefont {Silvestri}\ \emph {et~al.}(2014)\citenamefont
  {Silvestri}, \citenamefont {Jacobs}, \citenamefont {Dunjko},\ and\
  \citenamefont {Olshanii}}]{olshanii}%
  \BibitemOpen
  \bibfield  {author} {\bibinfo {author} {\bibfnamefont {L.}~\bibnamefont
  {Silvestri}}, \bibinfo {author} {\bibfnamefont {K.}~\bibnamefont {Jacobs}},
  \bibinfo {author} {\bibfnamefont {V.}~\bibnamefont {Dunjko}},\ and\ \bibinfo
  {author} {\bibfnamefont {M.}~\bibnamefont {Olshanii}},\ }\bibfield  {title}
  {\bibinfo {title} {Typical, finite baths as a means of exact simulation of
  open quantum systems},\ }\href@noop {} {\bibfield  {journal} {\bibinfo
  {journal} {Physical Review E}\ }\textbf {\bibinfo {volume} {89}},\ \bibinfo
  {pages} {042131} (\bibinfo {year} {2014})}\BibitemShut {NoStop}%
\bibitem [{\citenamefont {Lotshaw}\ and\ \citenamefont {Kellman}(2019)}]{shaw}%
  \BibitemOpen
  \bibfield  {author} {\bibinfo {author} {\bibfnamefont {P.~C.}\ \bibnamefont
  {Lotshaw}}\ and\ \bibinfo {author} {\bibfnamefont {M.~E.}\ \bibnamefont
  {Kellman}},\ }\bibfield  {title} {\bibinfo {title} {Simulating quantum
  thermodynamics of a finite system and bath with variable temperature},\
  }\href@noop {} {\bibfield  {journal} {\bibinfo  {journal} {Physical Review
  E}\ }\textbf {\bibinfo {volume} {100}},\ \bibinfo {pages} {042105} (\bibinfo
  {year} {2019})}\BibitemShut {NoStop}%
\bibitem [{\citenamefont {Faria}\ and\ \citenamefont
  {Bonan{\c{c}}a}(2020)}]{faria}%
  \BibitemOpen
  \bibfield  {author} {\bibinfo {author} {\bibfnamefont {A.~M.}\ \bibnamefont
  {Faria}}\ and\ \bibinfo {author} {\bibfnamefont {M.~V.}\ \bibnamefont
  {Bonan{\c{c}}a}},\ }\bibfield  {title} {\bibinfo {title} {Verification of
  finite bath fluctuation theorem for a non-ergodic system},\ }\href@noop {}
  {\bibfield  {journal} {\bibinfo  {journal} {Journal of Physics A:
  Mathematical and Theoretical}\ }\textbf {\bibinfo {volume} {53}},\ \bibinfo
  {pages} {345002} (\bibinfo {year} {2020})}\BibitemShut {NoStop}%
\bibitem [{\citenamefont {Riera-Campeny}\ \emph {et~al.}(2022)\citenamefont
  {Riera-Campeny}, \citenamefont {Sanpera},\ and\ \citenamefont
  {Strasberg}}]{riera}%
  \BibitemOpen
  \bibfield  {author} {\bibinfo {author} {\bibfnamefont {A.}~\bibnamefont
  {Riera-Campeny}}, \bibinfo {author} {\bibfnamefont {A.}~\bibnamefont
  {Sanpera}},\ and\ \bibinfo {author} {\bibfnamefont {P.}~\bibnamefont
  {Strasberg}},\ }\bibfield  {title} {\bibinfo {title} {Open quantum systems
  coupled to finite baths: A hierarchy of master equations},\ }\href@noop {}
  {\bibfield  {journal} {\bibinfo  {journal} {Physical Review E}\ }\textbf
  {\bibinfo {volume} {105}},\ \bibinfo {pages} {054119} (\bibinfo {year}
  {2022})}\BibitemShut {NoStop}%
\bibitem [{\citenamefont {Hakim}\ and\ \citenamefont
  {Ambegaokar}(1985)}]{hakim}%
  \BibitemOpen
  \bibfield  {author} {\bibinfo {author} {\bibfnamefont {V.}~\bibnamefont
  {Hakim}}\ and\ \bibinfo {author} {\bibfnamefont {V.}~\bibnamefont
  {Ambegaokar}},\ }\bibfield  {title} {\bibinfo {title} {Quantum theory of a
  free particle interacting with a linearly dissipative environment},\
  }\href@noop {} {\bibfield  {journal} {\bibinfo  {journal} {Physical Review
  A}\ }\textbf {\bibinfo {volume} {32}},\ \bibinfo {pages} {423} (\bibinfo
  {year} {1985})}\BibitemShut {NoStop}%
\bibitem [{\citenamefont {Lifshitz}\ and\ \citenamefont
  {Pitaevskii}(2013)}]{lp}%
  \BibitemOpen
  \bibfield  {author} {\bibinfo {author} {\bibfnamefont {E.~M.}\ \bibnamefont
  {Lifshitz}}\ and\ \bibinfo {author} {\bibfnamefont {L.~P.}\ \bibnamefont
  {Pitaevskii}},\ }\href@noop {} {\emph {\bibinfo {title} {Statistical physics:
  theory of the condensed state}}},\ Vol.~\bibinfo {volume} {9}\ (\bibinfo
  {publisher} {Elsevier},\ \bibinfo {year} {2013})\BibitemShut {NoStop}%
\bibitem [{\citenamefont {Bedeaux}\ and\ \citenamefont {Mazur}(1974)}]{mazur}%
  \BibitemOpen
  \bibfield  {author} {\bibinfo {author} {\bibfnamefont {D.}~\bibnamefont
  {Bedeaux}}\ and\ \bibinfo {author} {\bibfnamefont {P.}~\bibnamefont
  {Mazur}},\ }\bibfield  {title} {\bibinfo {title} {Brownian motion and
  fluctuating hydrodynamics},\ }\href@noop {} {\bibfield  {journal} {\bibinfo
  {journal} {Physica}\ }\textbf {\bibinfo {volume} {76}},\ \bibinfo {pages}
  {247} (\bibinfo {year} {1974})}\BibitemShut {NoStop}%
\end{thebibliography}%


\clearpage

\pagenumbering{arabic}

\widetext
\appendix

\renewcommand{\theequation}{A.\arabic{equation}}
\setcounter{equation}{0}


\section{Brownian stochastic motion and correlation functions}
\label{app-correlations}

We start by presenting the solution for the following Langevin equation with Ohmic friction and white noise [cf.~(\ref{lang}, \ref{oligo})]:
\begin{align}\label{langevin-11}
    \ddot{X}(t)&=~~b\dot{Y}(t) - \omega_0^2 X(t) - \gamma \dot{X}(t) + \xi_x(t) 
    \\
    \label{langevin-111}
    \ddot{Y}(t)&=-b\dot{X}(t) - \omega_0^2 Y(t) - \gamma \dot{Y}(t) + \xi_y(t)
    \\
    \la \xi_\alpha(t)\xi_\beta(t')\ra &= 2\gamma\,T \delta_{\alpha\beta} \delta(t-t') \qquad \alpha,\beta=x,y. \label{noise-corr-def}
\end{align}
Note that $X$ and $Y$ couple only via the magnetic field $b$; cf.~(\ref{mag}). The third equation in (\ref{lang}) is decoupled from (\ref{langevin-11}, \ref{langevin-111}). 

We solve (\ref{langevin-11}--\ref{noise-corr-def}) via the Laplace transform:
\begin{equation}\label{after-laplace}
    \begin{pmatrix}
    s^2+\omega_0^2+\gamma s & -bs \\
    bs & s^2+\omega_0^2+\gamma s \\
    \end{pmatrix}
    \begin{pmatrix}
    \hat{X}\\\hat{Y}
    \end{pmatrix}
    = 
    \begin{pmatrix}
    \hat{\xi}_x + X(0) (s+\gamma) + V_{x}(0) - b Y(0)
    \\
    \hat{\xi}_y + Y(0) (s+\gamma) + V_{y}(0) + b X(0)
    \end{pmatrix}.
\end{equation}
Inverting the matrix on the LHS amounts to transposing it and dividing it on its determinant. Hence, we get
\begin{align}\label{XY-laplase}
\hat{X}(s)=& X(0)\left(\hat{K}_{0}+b \hat{H}_{1}\right)
+Y(0)\left(\hat{K}_{1}-b \hat{H}_{0}\right) \\
&+V_{x}(0)\hat{H}_{0} +V_{y}(0)\hat{H}_{1}  +\hat{H}_{0} \hat{\xi}_{x}+\hat{H}_{1} \hat{\xi}_{y} 
\\
\hat{Y}(s)=& X(0)\left(-\hat{K}_{1}+H_{0} b\right)+Y(0)\left(\hat{K}_{0}+b \hat{H}_{1}\right) \\
&- V_{x}(0)\hat{H}_{1}+V_{y}(0)\hat{H}_{0}  -\hat{H}_{1} \hat{\xi}_{x}+\hat{H}_{0} \hat{\xi}_{y}
\end{align}
where
\begin{align}
{
\everymath={\displaystyle}
    \begin{array}{rlrl}
    \hat{H}_0&=\frac{s^2+\gamma \, s+\omega_0^2}{\Delta} 
    &\qquad
    \hat{H}_1&=\frac{b\,s}{\Delta}
    \\[10pt]
    \hat{K}_0&=\hat{H}_0 (s+\gamma) 
    &\qquad
    \hat{K}_1&=\hat{H}_1 (s+\gamma)
\end{array}
}
\end{align}
and the $\Delta$ is the determinant of the matrix in (\ref{after-laplace}):
\begin{align}
    \Delta&=(b\, s)^2+\left(s^2+\gamma\,  s+\omega_0^2\right)^2
    =\prod_{i=1}^4 (s-p_i)
\end{align}
where $p_i$ are roots of $\Delta=0$:
\begin{align} \label{poles}
    p_{1,2}=\frac{1}{2} \left(-\gamma+i b \pm\sqrt{(\gamma-ib)^2-4 \omega_0^2}\right) \qquad p_{3,4} = p_{1,2}^*
\end{align}
We also define residues of $\tfrac{1}{\Delta}$ which has 4 simple poles $\{p_i\}$ as 
\begin{equation}
    r_i=\frac{1}{\prod_{j\ne i} (p_i-p_j)} \qquad i,j=1,2,3,4
\end{equation}
thus, the inverse Laplace transform of the above kernels are (recall Cauchy's residue theorem)
\begin{align}\label{H01}
    H_0(t)= \sum_{i=1}^4 r_i (p_i^2+\gamma  p_i+\omega_0^2) e^{p_i t}
    \qquad
    H_1(t)= \sum_{i=1}^4 r_i b\, p_i \, e^{p_i t}
\end{align}
and similarly, we get analytical expressions for $K_0(t), K_1(t)$. Using these functions together with the convolution theorem for the Laplace transform, we get
\begin{align} \label{nolabl}
    X(t) =&X(0)\left(K_0(t)+bH_1(t)\right)
    +Y(0)\left(K_1(t)-bH_0(t)\right) 
    +V_x(0)H_0(t)+V_y(0)H_1(t) \\ \nonumber
    &+\int_0^t \d t' H_0 (t-t') \xi_x(t')
    +\int_0^t \d t' H_1 (t-t') \xi_y(t')
    \\ \label{Yt}
    Y(t) =&X(0)\left(-K_1(t)+bH_0(t)\right)
    +Y(0)\left(K_0(t)+bH_1(t)\right)
    -V_x(0)H_1(t)+V_y(0)H_0(t) \\ \nonumber
    &-\int_0^t \d t' H_1 (t-t') \xi_x(t')
    +\int_0^t \d t' H_0 (t-t') \xi_y(t')
\end{align}

We assume that distribution of random variables $X(0)$ and $Y(0)$ (initial values) are unbiased, symmetric and independent from each other:
\BEA
\label{copt}
{\cal P}(X(0), Y(0))={\cal P}(X(0)){\cal P}( Y(0)),\qquad {\cal P}(-x)={\cal P}(x),
\EEA
which in particular implies:
\begin{align}
    \label{initial-XY}
    &\la X(0)\ra =\la Y(0)\ra=0,\\
    &\la X(0)^2 \ra =\la Y(0)^2\ra\equiv\la X_0^2\ra, \\
    &\la V_x(0)^2 \ra =\la V_y(0)^2\ra\equiv\la V_0^2\ra, \\
    &\la X(0)Y(0)\ra=\la V_x(0)V_y(0)\ra=0. 
\end{align}
Also note that, $X(0)$, $Y(0)$, $\xi_x(t\geq 0)$ and $\xi_y(t\geq 0)$ are independent random variables due to the assumed system-bath initial state. 
Then the correlation function $C_{XY}(t,t')=\la X(t) Y(t')\ra$ will be
\begin{align}
    C_{XY}(t,s)=& \la V_0^2\ra\left( H_{1}(t) H_{0}(s) -H_{0}(t) H_{1}(s)\right)
    \\
    &+\la X_0^2\ra\Big( \left(K_{0}(t)+b H_{1}(t)\right)\left(-K_{1}(s)+b H_{0}(s)\right) 
    \\
    &\qquad\qquad+\left(K_{1}(t)-b H_{0}(t)\right)\left(K_{0}(s)+b H_{1}(s)\right)\Big)
    \\
    &+\int_{0}^{t} \int_{0}^{s}\d t'\d s' \la\xi\left(t^{\prime}\right) \xi(s')\ra
    \left(-H_{0}(t-t') H_{1}(s-s') 
    +H_{0}(s-s') H_{1}(t-t')\right)
\end{align}
where the last line can be simplified as noise is white:
\begin{align}
\label{Cxy-def}
    C_{XY}(t,s)=& \la V_0^2\ra\left( H_{1}(t) H_{0}(s) -H_{0}(t) H_{1}(s)\right)
    \\
    &+\la X_0^2\ra\Big( \left(K_{0}(t)+b H_{1}(t)\right)\left(-K_{1}(s)+b H_{0}(s)\right) 
    \\
    &\qquad\qquad+\left(K_{1}(t)-b H_{0}(t)\right)\left(K_{0}(s)+b H_{1}(s)\right)\Big)
    \\
    &+2\gamma T\int_{0}^{\text{min}(t,s)} \d y 
    \left(-H_{0}(t-y) H_{1}(s-y) 
    +H_{0}(s-y) H_{1}(t-y)\right).
\end{align}
Note that $C_{XY}$ is antisymmetric:
\BEA
C_{XY}(t,s)=-C_{XY}(s,t).
\label{cooper}
\EEA

Similarly, the correlation function $C_{XX}(t,s)=\la X(t) X(s)\ra$ will be
\begin{align}
\label{Cxx-def}
    C_{XX}(t,s)=& \la V_0^2\ra\left( H_{0}(t) H_{0}(s) +H_{1}(t) H_{1}(s)\right)
    \\
    &+\la X_0^2\ra\Big( \left(K_{0}(t)+b H_{1}(t)\right)\left(K_{0}(s)+b H_{1}(s)\right) 
    \\
    &\qquad\qquad+\left(K_{1}(t)-b H_{0}(t)\right)\left(K_{1}(s)-b H_{0}(s)\right)\Big)
    \\
    &+2\gamma T\int_{0}^{\text{min}(t,s)} \d y 
    \left(H_{0}(t-y) H_{0}(s-y) 
    +H_{1}(t-y) H_{1}(s-y)\right)
\end{align}

Therefore, the position-velocity autocorrelation functions can be derived by:
\begin{align}
    C_{X U_x}(t,s) \equiv \la X(t) U_x(s)\ra &= \frac{\d}{\d s}\la X(t) X(s)\ra = \frac{\d}{\d s} C_{XX}(t,s)
    \\
    C_{X U_y}(t,s) \equiv \la X(t) U_y(s)\ra &= \frac{\d}{\d s}\la X(t) Y(s)\ra = \frac{\d}{\d s} C_{XY}(t,s) \label{Cxvy-def}
\end{align}

\section{Equations of motion and Langevin equations }
\label{app-langevin}

Recall the full Lagrangian $\mathcal{L}_{S}+\mathcal{L}_{B}$ of our model [cf.~(\ref{lagr_sys}, \ref{lagr_bath})]:
\begin{align}
\mathcal{L}=&~\frac{1}{2}\dot{\R}^2 -\frac{\omega_0^2}{2}\R^2+\A\R\nonumber \\
&+\sum_{k=1}^N \left[\frac{m_k}{2}\dot{\r}_k^2 -
\frac{m_{k}\omega_{k}^{2}}{2} \left(\r_{k}-\frac{c_k\R}{m_k\omega_k^2}\right)^{2}\right]
\label{lagr_full}
\end{align}
with $\R=(X,Y)$, $\r_k=(x_k, y_k)$ and the vector potential for the stationary and homogeneous magnetic field reads $\A=(A_x,A_y)=(-\frac{bY}{2}, \frac{bX}{2})$; cf.~(\ref{lagr_sys}--\ref{lagr_bath}). For simplicity we assumed that the motion is two-dimensional [cf.~(\ref{langevin-11}, \ref{langevin-111})], i.e. we did not account explicitly for the third coordinate of the bath, since for the harmonic external potential of the Brownian particle (central oscillator), the third coordinates decouple from the other two coordinates. 

From Lagrange equations we get equations of motion for the Brownian particle:
\begin{align}
    \label{E-LX}
    \ddot{X}&=b\dot{Y} - \omega_0^2 X+\sum_k c_k \left(x_k-X\frac{c_k}{m_k \omega_k^2}\right),  \\
    \label{E-LY}
    \ddot{Y}&=-b\dot{X} - \omega_0^2 Y + \sum_k c_k \left(y_k-Y\frac{c_k}{m_k \omega_k^2}\right),
\end{align}
and for bath oscillators:
\begin{align}
\label{motion-x}
    m_k \ddot{x}_k + m_k \omega_k^2 x_k &= c_k X(t) 
    \\
    m_k \ddot{y}_k + m_k \omega_k^2 y_k &= c_k Y(t)
    \label{motion-y}
\end{align}
We can find exact solutions for $x_k(t), y_k(t)$ in (\ref{motion-x}, \ref{motion-y}) assuming that $X(t), Y(t)$ are given. Employ the Laplace transform in (\ref{motion-x}):
\begin{align}
    \hat{x}_{k}(s)=v_{xk}(0)\frac{1}{s^{2}+\omega_{k}^{2}}+x_{k}(0)\frac{s }{s^{2}+\omega_{k}^{2}}+\frac{c_k}{m_{k}\left(s^{2}+\omega_{k}^{2}\right)} \hat{X}(s).
\end{align}
Then the inverse Laplace transform produces:
\begin{align}\label{osc-sol-1}
    x_k(t)&=x_k(0) \cos(\omega_k t) + \frac{v_{xk}(0)}{\omega_k} \sin(\omega_k t)+\frac{c_k}{m_k\omega_k} \int_0^t \d t' \sin(\omega_k (t-t')) X(t').
\end{align}
A similar formula holds for $y(t)$. Now integrate by parts the final integral in (\ref{osc-sol-1}):
\begin{align}
    \label{osc-sol-2}
    x_k(t)-X(t)\frac{ c_k}{m_k \omega_k^2}=&\left(x_k(0)-X(0)\frac{ c_k}{m_k \omega_k^2}\right) \cos(\omega_k t)+ \frac{v_{xk}(0)}{\omega_k} \sin(\omega_k t)
    \\
    &-\frac{c_k}{m_k\omega_k^2} \int_0^t \d t' \cos(\omega_k (t-t')) V_x(t'),
\end{align}
and insert these results back to (\ref{E-LX}, \ref{E-LY})
\begin{align}
\label{langevin-1}
    \ddot{X}&=b\dot{Y} - \omega_0^2 X - r_x(t) + \xi_x(t), 
    \\
    \ddot{Y}&=-b\dot{X} - \omega_0^2 Y - r_y(t) + \xi_y(t).
    \label{langevin-2}
\end{align}
Eqs.~(\ref{langevin-1}, \ref{langevin-2}) are Langevin equations for the Brownian particle, where we defined for friction $r_s$ and noise $\xi_s$:
\begin{align}
\label{friction-def}
    r_s(t) &= \int_0^t \d t' \zeta(t-t') V_s(t'),\qquad s=x,y 
    \\
    \zeta(t)&=\sum_k \frac{c_k^2}{m_k \omega_k^2} \cos(\omega_k t)
    \\ \label{noise-x-def}
    \xi_x(t) &= \sum_k c_k\overline{x_k} \cos(\omega_k t)+ 
    \frac{c_k}{\omega_k} \overline{ v_{xk}} \sin(\omega_k t),
    \\
    \xi_y(t) &= \sum_k c_k\overline{y_k} \cos(\omega_k t)+ 
    \frac{c_k}{\omega_k}\overline{v_{yk}} \sin(\omega_k t), 
    \label{noise-y-def}
\end{align}
where we define
\begin{align}
    \left(x_k(0)-X(0)\frac{ c_k}{m_k \omega_k^2}\right)=\overline{x_{k}},
    \qquad 
    &v_{xk}(0)=\overline{v_{xk}},\qquad \label{xbardef}
    \\
    \left(y_k(0)-Y(0)\frac{ c_k}{m_k \omega_k^2}\right)=\overline{y_{k}},
    \qquad
    &v_{yk}(0)=\overline{v_{yk}}.
\end{align}

\subsection{Equilibrium bath}
If we take Gibbs distribution for initial conditions for oscillators, i.e.
\begin{align}
    \label{bath-gibbs}
    (x_k(0), y_k(0), v_{xk}(0), v_{yk}(0))\sim \mathcal{N}e^{-H/T}
\end{align}
where $\mathcal{N}$ normalisation constant and
\begin{align}
    H=\sum_k&\left[\frac{1}{2}m_k \omega_k^2\;\ol{y_k}^2 
    +\frac{1}{2}m_k \omega_k^2\;\ol{y_k}^2  +\frac{1}{2}m_k \;\ol{v_{xk}}^2+\frac{1}{2}m_k \;\ol{v_{yk}}^2\right]
\end{align}
we get
\begin{align}
    \la \xi_\alpha(t)\xi_\beta(t')\ra = T \delta_{\alpha\beta} \zeta(t-t') \qquad \alpha,\beta=x,y \label{noise-corr-def-2}
\end{align}
We see that $\zeta(t)$ appears both in friction memory (\ref{friction-def}) and noise correlation (\ref{noise-corr-def-2}), so this implies  fluctuation–dissipation relation holds.

\subsection{Continuous spectrum of oscillators}
Now we consider some specific bath, which has uniformly spaced frequencies with Drude-Ullersma's spectrum:
\begin{align}
    \omega_n=\delta\omega \; n \qquad n=1,2,3,\dots
    \\
    \label{c-spectr}
    c_n = \sqrt{\frac{2\gamma m_n \omega_n^2 \delta\omega}{\pi}\frac{\theta^2}{\omega_n^2+\theta^2}} 
\end{align}
we are going to take continuum limit where the spacing of frequencies $\delta\omega\rightarrow0$. The memory kernel (which is also the noise correlation function) will be:
\begin{align}
    \zeta(t)=& \sum_{n=1}^\infty \frac{2\gamma \,\delta\omega}{\pi} \cos(\omega_n t ) \frac{\theta^2}{\omega_n^2+\theta^2}
    \\
    \rightarrow& \int_0^\infty 
    \frac{2\gamma}{\pi} \cos(\omega t ) \frac{\theta^2}{\omega^2+\theta^2} \d \omega \qquad\text{as} ~~\delta\omega\rightarrow0
    \\
    =&\gamma \;\theta e^{-\theta t} 
\end{align}
This memory function represents memory time $1/\theta$. In the limit $\theta\rightarrow\infty$ we are going to have memoryless/Ohmic friction and white noise:
\begin{align}
    \zeta(t) =& 2 \gamma\,\delta(t) \label{white-zeta}
    \\
    c(\omega)=&\sqrt{\frac{2\gamma \omega^2}{\pi}}
    \\
    m\equiv& 1
    \label{white-zeta_1}
\end{align}
here we took $m=1$ for all the oscillators; cf.~(\ref{oligo}).

\subsection{Symmetries of the model}\label{app-symmetries}

We have seen that by the initial bath distribution (\ref{bath-gibbs}) and the spectrum of the oscillator couplings (\ref{c-spectr}) we recover the same Langevin equation (\ref{langevin-11}-\ref{noise-corr-def}). Note that these Langevin equations are invariant under rotation in $X-Y$ plane. Additionally, the initial distribution of $X(0)$, $Y(0)$, $V_x(0)$, $V_y(0)$ is invariant under $90^\circ$ rotations; see (\ref{initial-XY}). Thus, the solution of these equations must also be invariant under $90^\circ$ rotations. More specifically, for any function $f(X,V_x, Y, V_y)$ we have
\begin{equation}
    \la f(X(t), V_x(t), Y(s), V_y(s)\ra = \la f(Y(t), V_y(t), -X(s), -V_x(s)) \ra
\end{equation}
which corresponds to the $90^\circ$ rotation. Eq.~(\ref{cooper}) is a consequence of it.

Similarly, the Lagrangian (\ref{lagr_full}) and initial distribution (\ref{bath-gibbs}, \ref{initial-XY}) are rotation symmetric, and for any function $f\left(X, V_x, \{x_k, v_{xk}\}, Y,V_y, \{y_k, v_{yk}\}\right)$  with $\{x_k, v_{xk}\}\equiv\{x_1, x_2, \dots,v_{x1}, v_{x2}, \dots\}$ we have
\begin{align}
    &\la f\left(X(t), V_x(t), \{x_k(t), v_{xk}(t)\}, Y(s),V_y(s), \{y_k(s), v_{yk}(s)\} \right) \ra
    \\
    &\qquad=\la f\left(Y(t), V_y(t), \{y_k(t), v_{yk}(t)\}, -X(s),-V_x(s), \{-x_k(s), -v_{xk}(s)\}\right) \ra
\end{align}
Again, this follows from $90^\circ$  rotation. Or in other words, when we take the average over the distribution of the solution, we can replace $X\rightarrow Y$ and $Y\rightarrow -X$ and the time derivatives accordingly. For example
\BEA
    \label{symm-xvy}
    &\la x_k(t) v_{yk}(t)-y_k(t) v_{xk}(t)\ra = 2 \la x_k(t)\, v_{yk}(t)\ra
    \\
    \label{symm-Xy}
    &\la X(t) v_{yk}(s) \ra = -\la Y(t) v_{xk}(s) \ra
    \\
    \label{symm-Xvy}
    &\la X(t) \overline{y_{k}} \ra = -\la Y(t) \overline{x_k} \ra
    \\
    \label{symm-etc}
    &\la X^2 \ra = \la Y^2 \ra, \quad \la X x_k \ra = \la Y y_k \ra, \quad \la v_{xk}^2 \ra = \la v_{yk}^2 \ra \quad \text{etc.}
\EEA
    
\subsection{Linear momentum of bath modes}\label{app-momentum}

Starting from equations of motion generated by (\ref{lagr_full}) [see also (\ref{lagr_sys}, \ref{lagr_bath})] we can derive for $\omega_0$ and ${\bf B}={\bf A}=0$|i.e. no external potential and no magnetic field for the Brownian particle|the following conservation law for the linear momentum:
\BEA
\frac{\d }{\d t}\left[\sum_k
\frac{c_k}{\omega_k^2}\, \dot\r_k+\dot\R
\right]=0,
\label{gnu}
\EEA
where we note that only a part of the mode linear momentum $m_k\dot\r_k$ participates in the conservation law; $m_k\dot\r_k$ is obviously not conserved, since each mode feels an external potential. We now employ (\ref{osc-sol-2}) with initial conditions [cf.~(\ref{ino})]:
\BEA
\label{kho1}
&&\langle\r_k(0) \rangle=\langle\dot{\r}_k(0)
\rangle=\langle\R(0) \rangle=0,\\
&&\langle\dot{\R}(0) \rangle\not=0,
\label{kho2}
\EEA
and also use the Langevin equation that under (\ref{kho1}, \ref{kho2}) and (\ref{white-zeta}, \ref{white-zeta_1}) reads
\BEA
\frac{\d}{\d t}\langle \V\rangle=-\gamma
\langle \V\rangle, \qquad 
\langle \V(t)\rangle=e^{-\gamma t}
\langle \V(0)\rangle,\qquad \V(t)=\dot{\R}(t).
\label{ansar}
\EEA
Recall that for a free Brownian particle the thermalization according to the Langevin equation is incomplete: the velocity distribution relaxes towards Maxwell's density (in particular, $\langle \V(t)\rangle$ goes to zero according to (\ref{ansar})), while the coordinate does not relax, since it makes a free Brownian motion. 

We find
\BEA
\label{guli0}
\langle\dot{\r}_k(t) \rangle&=&\frac{c_k}{m_k\omega_k}
\int_0^t\d s\, \sin[\omega_k s]
\langle\V(t-s) \rangle \\
&=&\frac{c_k \langle\V(0) \rangle}{m_k\omega_k}
\left[
\frac{\gamma \sin[\omega_k t]-\omega_k \cos[\omega_k t]}{\gamma^2+\omega_k^2}+\frac{\omega_k e^{-\gamma t}}{\gamma^2+\omega_k^2}
\right].
\label{guli}
\EEA
Hence we obtain from (\ref{guli}):
\BEA
\sum_k\frac{c_k}{\omega_k^2}\langle\dot{\r}_k(t) \rangle=\frac{2\gamma}{\pi}\langle\V(0) \rangle
\int_0^\infty \d\omega\, 
\left[
\frac{\frac{\gamma}{\omega} 
\sin[\omega t]- \cos[\omega t]}{\gamma^2+\omega^2}+\frac{ e^{-\gamma t}}{\gamma^2+\omega^2}
\right],
\label{olu}
\EEA
where employing [$t>0$ and $\gamma>0$]
\BEA
\int_{-\infty}^\infty \d\omega\, 
\frac{\cos[\omega t]}{\gamma^2+\omega^2}
=\frac{ \pi e^{-\gamma t} }{\gamma},\qquad 
\int_{-\infty}^\infty \d\omega\, 
\frac{ \frac{\gamma}{\omega} \sin[\omega t]}{\gamma^2+\omega^2}=\frac{\pi}{\gamma}(1-e^{-\gamma t}),
\EEA
we confirm from (\ref{olu}) that the conservation (\ref{gnu}) indeed holds. 

Returning to (\ref{guli}) we see that the mean linear momentum of each mode oscillates in time. This fact contrasts the mean angular momentum of each mode that (under a non-zero magnetic field) relaxes in time to a well-defined, frequency-dependent value. Another pertinent difference is that a non-zero linear momentum for bath modes is achieved only due to $\langle\V(0) \rangle\not=0$, i.e. the Brownian particle should have a non-zero initial momentum for transferring it to the bath modes according to the conservation law (\ref{gnu}). Such a direct relation need not hold for the angular momentum, where the bath modes acquire angular momentum also without the initial angular momentum of the particle. 

Note that generic finite collective observables of the bath lose the non-zero linear momentum in the long-time limit. Indeed, we note from (\ref{guli}):
\BEA
\sum_{\omega_1}^{\omega_2}\frac{c_k}{\omega_k^2}\langle\dot{\r}_k(t) \rangle=\frac{2\gamma}{\pi}\langle\V(0) \rangle
\int_{\omega_1}^{\omega_2} \d\omega\, 
\left[
\frac{\frac{\gamma}{\omega} 
\sin[\omega t]- \cos[\omega t]}{\gamma^2+\omega^2}+\frac{\omega_k e^{-\gamma t}}{\gamma^2+\omega^2}
\right],
\label{olun}
\EEA
where $\omega_1<\omega_2$ are finite frequencies that define the collective observable.
For $t\gg 1/\gamma$ the contribution $O(e^{-\gamma t})$ in (\ref{olun}) can be neglected. The remaining two integrals converge to zero as $O(1/t)$ for long times $t$ and for finite and non-zero values of $\omega_1$ and $\omega_2$:
\BEA
\label{ho1}
\int_{\omega_1}^{\omega_2} \d\omega\, 
\frac{\frac{\gamma}{\omega} 
\sin[\omega t] }{\gamma^2+\omega^2}\simeq 
\frac{1}{\gamma} \int_{\omega_1 t}^{\omega_2 t} \d\omega\, 
\frac{\sin[\omega ] }{\omega}=\frac{1}{\gamma}\, O(\frac{1}{t}), \\
\int_{\omega_1}^{\omega_2} \d\omega
\frac{ \cos[\omega t]}{\gamma^2+\omega^2}\simeq \frac{\sin[\omega_2 t]-\sin[\omega_1 t]}{t\gamma^2},
\label{ho2}
\EEA
where we note that the last relation in (\ref{ho1}) does not apply for $\omega_1=0$. At that specific (non-generic) value of $\omega_1$ the integral in (\ref{ho1}) converges to a finite value for long-times, since $\int_{0}^{\infty} \d\omega\, \frac{\sin[\omega] }{\omega}=\frac{\pi}{2}$.

Altogether, we conclude that for the Brownian particle and finite observables of the bath the 
linear momentum is dissipated away. This discussion can be conveniently summarized as follows:
If we define linear momentum density over the modes at time $t$:
\begin{align}
    {\bf \Pi}(\omega;\, t) = 
    \langle \V(0)\rangle \frac{2\gamma}{\pi}
\left[
\frac{\frac{\gamma}{\omega} 
\sin[\omega t]- \cos[\omega t]}{\gamma^2+\omega^2}+\frac{ e^{-\gamma t}}{\gamma^2+\omega^2}
\right],
\end{align}
then for $t\rightarrow\infty$ this quantity weakly converges to 
\begin{equation}
    {\bf \Pi}(\omega;\, t\rightarrow\infty) =2 \langle\V(0)\rangle \delta(\omega), \qquad \int_0^\infty \d \omega
    {\bf \Pi}(\omega;\, t\rightarrow\infty)=\langle\V(0)\rangle.
\end{equation}

For completeness, let us also consider a situation where (\ref{kho1}) holds, but instead of (\ref{kho2}) we take $\langle \V(0)\rangle=0$. However, now there is a constant and homogeneous electric field ${\bf E}$ that is acting on the Brownian particle. Hence the Langevin equation reads instead of (\ref{ansar}):
\BEA
\frac{\d}{\d t}\langle \V\rangle=-\gamma
\langle \V\rangle+{\bf E}, \qquad 
\langle \V(t)\rangle=\frac{{\bf E}}{\gamma }(1-e^{-\gamma t}),
\label{ansar2}
\EEA
where we assumed a unit charge. Note that this situation is different from that of (\ref{ansar}), since now there is no thermalization even for the velocity of the Brownian particle, because now this particle moves for long times with a constant velocity ${\bf E}/\gamma$; cf.~(\ref{ansar2}). Also, the conservation law (\ref{gnu}) changes to 
\BEA
\frac{\d }{\d t}\left[\sum_k
\frac{c_k}{\omega_k^2}\, \dot\r_k+\dot\R-{\bf E}\,t
\right]=0,
\label{gnu2}
\EEA
which means that the overall momentum $\sum_k
\frac{c_k}{\omega_k^2}\, \dot\r_k+\dot\R$ increases monotonously in time. 

Now (\ref{guli0}) still holds and putting there (\ref{ansar2}) we find
\BEA
\label{guli3}
\langle\dot{\r}_k(t) \rangle&=&\frac{c_k}{m_k\omega_k}\,\frac{{\bf E}}{\gamma}\,\frac{1-\cos[\omega_k t]}{\omega_k} \\
&-&\frac{{\bf E}}{\gamma}\,\frac{c_k}{m_k\omega_k}
\left[
\frac{\gamma \sin[\omega_k t]-\omega_k \cos[\omega_k t]}{\gamma^2+\omega_k^2}+\frac{\omega_k e^{-\gamma t}}{\gamma^2+\omega_k^2}
\right].
\label{guli4}
\EEA
The contribution coming from (\ref{guli4}) was studied by us above, hence we focus on (\ref{guli3}). Note that 
\BEA
\sum_k \frac{c_k}{\omega_k^2}\, \langle\dot\r_k\rangle=\frac{2{\bf E}}{\pi}
\int_{0}^{\infty} \d\omega\, 
\frac{1-\cos[\omega t] }{\omega^2}=t\,{\bf E}.
\EEA
This relation validates the conservation law (\ref{gnu2}). Now for $\omega_2>\omega_1>0$ we find from (\ref{guli3}):
\BEA
\sum_{\omega_1}^{\omega_2} \frac{c_k}{\omega_k^2}\, \langle \dot\r_k\rangle =\frac{2{\bf E}}{\pi}
\int_{\omega_1}^{\omega_2} \d\omega\, 
\frac{1-\cos[\omega t] }{\omega^2}, 
\EEA
which converges for $t\to\infty$ to a finite positive value (for $\omega_2>\omega_1>0$):
\BEA
\sum_{\omega_1}^{\omega_2} \frac{c_k}{\omega_k^2}\, \langle \dot\r_k\rangle\to
\frac{2{\bf E}}{\pi}\left[
\frac{1}{\omega_1}-\frac{1}{\omega_2}
\right].
\label{ansar3}
\EEA
After (\ref{ho2}) we concluded that for the Brownian particle and finite observables of the bath the linear momentum is dissipated away. Here the situation is different: the linear momentum cannot be dissipated away, since for long times $t$ the Brownian particle and finite bath observables move with a constant linear momentum $\propto {\bf E}$; cf.~(\ref{ansar2}, \ref{ansar3}). The explanation of this is straightforward: at long times the particle moves with a constant linear momentum and drags along the bath modes. Instead, the acceleration induced by ${\bf E}$ is dissipated away from the particle and finite bath observables. 
A detailed comparison between the linear momentum and angular momentum is presented in Table~\ref{tabu}.

\begin{center}
\begin{table}
\caption{Comparison between the linear momentum and angular momentum is convoluted. Hence we separated 4 cases depending on the following 3 factors. {\it (i)} Initial conditions for the Brownian particle, where $\langle{\bf V}(0)\rangle$ and $\langle{\bf L}(0)\rangle$ mean the initial
velocity and angular momentum, respectively. {\it (ii)} External fields electric ${\bf E}$ and magnetic fields ${\bf B}$. Both are constant and homogeneous in space and only one of them is present for each case. {\it (iii)} The external harmonic potential with magnitude $\omega_0$ for the particle. For each of these 4 cases we indicate
whether and which conservation law is present, whether and to which extent the particle thermalizes for long times, and how collective bath observable behave. It is seen that the most reasonable comparison is to be done between the second and fifth columns, and this is the road we followed in the main text.  }
\begin{tabular}{ |c| c| c | c| c|}
\hline
       &  $\omega_0=0$ & $\omega_0=0$ & $\omega_0\not=0$ & $\omega_0\not=0$ \\
& ${\bf E}=0$ & ${\bf E}\not=0$ & ${\bf B}=0$ & ${\bf B}\not=0$ \\
& ${\bf B}=0$ & ${\bf B}=0$     & ${\bf E}=0$ & ${\bf E}=0$ \\
& $\langle{\bf V}(0)\rangle\not=0$ & $\langle{\bf V}(0)\rangle=0$ &
$\langle{\bf L}(0)\rangle\not=0$ & $\langle{\bf L}(0)\rangle=0$
\\ \hline
 Conservation law & yes, effective             & no & yes                      & yes, effective  \\  
                  & linear momentum            &    & orbital momentum         & orbital momentum \\  \hline
 Thermalization of particle & yes, for velocity              & no & yes & yes \\  
                            & distribution only              & do not forget ${\bf E}$ &  & \BvL theorem  \\  \hline
 Collective bath observables & forget $\langle{\bf V}(0)\rangle$  & do not forget ${\bf E}$ 
                             & forget $\langle{\bf L}(0)\rangle$  & do not forget ${\bf B}$ \\  \hline
\end{tabular}
\label{tabu}
\end{table}
\end{center}

\section{Angular momentum of a bath oscillator}
\label{app-angular-momentum}

Here we consider a specific oscillator from  (\ref{osc-sol-1}) and (\ref{xbardef}):
\begin{align}\label{x-sol}
    x(t)=&\ol{x} \,\cos(\omega t)
    +X(0) \frac{c}{m\omega^2} \cos(\omega t)
    + \frac{\ol{v_{x}}}{\omega} \sin(\omega t)
    \\
    &+\frac{c}{m\omega} \int_0^t \d t' \sin(\omega (t-t')) X(t')
\end{align}
where we dropped subscript $k$'s for simplicity. For the velocity we differentiate (\ref{x-sol}) (written in terms of $y$)
\begin{align}\label{v-sol}
    v_y(t)=&-\ol{y}\,\omega \sin(\omega t)
    -Y(0) \frac{c}{m\omega} \sin(\omega t) 
    + \ol{v_{y}}\, \cos(\omega t)
    \\
    &+\frac{c}{m} \int_0^t \d t' \cos(\omega (t-t')) Y(t')
\end{align}
Eventually, we want to calculate the correlation $\la x v_y - y v_x\ra$ at time $t$. But as the system is rotation symmetric, it is enough to consider $\la x v_y \ra$; see (\ref{symm-xvy}).
\begin{align}\label{xvy-v1-l1}
    \la x v_y\ra_t=&-\frac{c}{m} \int_0^t \la X(t')\ol{y}\ra \cos(\omega t')\d t'
    \\
    \label{xvy-v1-l2}
    &-\frac{c}{m\omega}  \int_0^t \la X(t')\ol{v_y}\ra \sin(\omega t') \d t'
    \\
    &- \frac{c^2}{m^2 \omega^2}  \int_0^t \la X(t) Y(0)\ra \cos(\omega t') \d t'
    \\
    \label{xvy-v1-l4}
    &+\frac{c^2}{m^2 \omega} \int_0^t\d t' \int_0^t \d t'' \la X(t')Y(t'')\ra \sin(\omega(t-t'))\cos(\omega(t-t''))
\end{align}
where we used the relations (\ref{symm-Xy}, \ref{symm-xvy}) following from rotation symmetry and the independence of the initial conditions (\ref{bath-gibbs}, \ref{initial-XY}).

We can calculate $\la X(t) \left(y(0)-\frac{Y(0) c}{m\omega^2}\right)\ra$ and similar terms using solutions for $X$ and $Y$ (\ref{nolabl}, \ref{Yt}). In (\ref{nolabl}) only $\xi_y(t)$ depends on $\overline{y}\equiv\left(y(0)-\frac{Y(0) c}{m\omega^2}\right)$ and $v_y$; i.e. the initial state of the bath, see (\ref{noise-y-def}). So we get

\begin{align}
    \la X(t) \ol{y}\ra
    =& T\frac{c}{m\omega^2} \int_0^t \d t' H_1(t-t') \cos(\omega t')
    \\
    \la X(t) \ol{v_y}\ra
    =& T \frac{c}{m\omega} \int_0^t \d t' H_1(t-t') \sin(\omega t')
\end{align}
Now we can simplify first two lines of  (\ref{xvy-v1-l1}, \ref{xvy-v1-l2}):
\begin{align}
    &-T\frac{c^2}{m^2\omega^2} \int_0^t \d t' \int_0^{t'} \d t'' H_1(t'-t'')
    \left(\cos(\omega t') \cos(\omega t'')+\sin(\omega t') \sin(\omega t'') \right)
    \\
    &\quad=-T\frac{c^2}{2m^2\omega^2} \int_0^t \int_0^{t} \d t' \d t'' H_1\left(|t'-t''|\right)\cos(\omega(t'-t''))
    \\
    &\quad=-T\frac{c^2}{m^2\omega^2} \int_0^t \d t' H_1\left(t'\right)\,(t-t')\cos(\omega t') 
\end{align}
Eq.~(\ref{xvy-v1-l4}) can also be simplified via (\ref{cooper}):
\begin{align}
    &\frac{c^2}{m^2 \omega} \int_0^t\d t' \int_0^t \d t'' \la X(t')Y(t'')\ra \frac{1}{2}\left(\sin\left(2\omega t-\omega(t'+t''))\right)+\sin(\omega(t''-t'))\right)
    \\
    &\quad=\frac{c^2}{m^2 \omega} \int_0^t \d t' \int_0^{t'}  \d t'' \la X(t')Y(t'')\ra\sin(\omega(t''-t'))
\end{align}

Putting it all together we get a simplified result for $\la xv_y \ra$:
\begin{align}\label{xvy-v2}
    \la x v_y\ra_t=&-T\frac{c^2}{m^2\omega^2} \int_0^t \d t' H_1\left(t'\right)\,(t-t')\cos(\omega t') 
    \\
    &- \frac{c^2}{m^2 \omega^2}  \int_0^t C_{XY}(t',0) \cos(\omega t') \d t'
    \\
    &-\frac{c^2}{m^2 \omega} \int_0^t \d t' \int_0^{t'}  \d t'' C_{XY}(t',t'') \sin(\omega(t'-t''))
\end{align}
and the expressions for $C_{XY}(t',t'')$ and $H_1(t')$ are given in  (\ref{H01}, \ref{Cxy-def}). 
Taking the limit $t\rightarrow\infty$ we get \footnote{\label{foot}
The expression that we are interested in is expressed only by the Laplace kernels (\ref{H01}) and its integrals w.r.t. time. These kernels are exponential functions of time and the integrals can be easily evaluated. However, the expression contains many exponential terms, the exact result is found using \textit{Mathematica}.}

\begin{align}
    L=2\la x v_y\ra_{t\rightarrow\infty}=&\delta\omega\,\frac{4\gamma b T}{\pi} \frac{(\omega-\omega_0)(\omega+\omega_0)\left((1-\sigma_V) \omega^{2}+(1-\sigma_X) \omega_0^{2}\right)}{b^{4} \omega^{4}+\left(\gamma^{2} \omega^{2}+\left(\omega^{2}-\omega_0^{2}\right)^{2}\right)^{2}+2 b^{2}\left(\gamma^{2} \omega^{4}-\left(\omega^{3}-\omega \omega_0^{2}\right)^{2}\right)}
    \\
    =&\delta\omega\,\frac{4\gamma b T}{\pi} \frac{(\omega^2-\omega_0^2)\left((1-\sigma_V) \omega^{2}+(1-\sigma_X) \omega_0^{2}\right)}{\left(\left(\omega^{2}-\omega_0^{2}+i \gamma \omega\right)^{2}-b^{2} \omega^{2}\right)\left(\left(\omega^{2}-\omega_0^{2}-i\gamma \omega\right)^{2}-b^{2} \omega^{2}\right)}
    \\
    =&\delta\omega\,\frac{4\gamma b T}{\pi} \frac{(\omega^2-\omega_0^2)\left((1-\sigma_V) \omega^{2}+(1-\sigma_X) \omega_0^{2}\right)}
    {\left((\omega^{2}-\omega_0^{2}-b\omega)^2+\gamma^2\omega^2 
    \right)
    \left((\omega^{2}-\omega_0^{2}+b\omega)^2+\gamma^2\omega^2\right)}
\end{align}
where 
\begin{align}\label{sigma-def}
    \la X(0)^2\ra=\la Y(0)^2\ra = \sigma_X \; \frac{T}{\omega_0^2}\quad\text{and}\quad   \la V_x(0)^2\ra =\la V_y(0)^2\ra = \sigma_V \; T
\end{align}

\section{Energy of a bath oscillator}
\label{app-energy}

From the Lagrangian (\ref{lagr_full}) we can see that conserved energy is
\begin{align}
    E_{\text{tot}}=&\frac{1}{2}\left(\dot{X}^2+\dot{Y}^2\right)+\frac{1}{2}\omega_0^2\left(X^2+Y^2\right)
    \\
    &-\sum_{k} c_k \left(X x_{k}+Y y_k\right) 
    \\
    &+\sum_{k} \frac{1}{2} m_{k} \left(\dot{x}_{k}^{2}+\dot{y}_{k}^{2}\right)+\frac{1}{2} m_{k} \omega_{k}^{2} \left(x_{k}^{2}+y_{k}^2\right)
    \\
    &+\left(X^{2}+Y^{2}\right) \sum_{k} \frac{1}{2} \frac{c_k^{2}}{m_{k} \omega_{k}^{2}}
\end{align}
Our concern is the average energy of the oscillator (again we drop subscript $k$):
\begin{align}
    E=& \frac{1}{2}m \la\dot{x}^{2}+\dot{y}^{2}\ra+\frac{1}{2} m \omega^{2} \la x^{2}+y^2\ra
    \\
    &+\la X^{2}+Y^{2}\ra  \frac{1}{2} \frac{c^{2}}{m \omega^{2}}
    - c \la X x+Y y\ra 
\end{align}
The system is rotation invariant, see (\ref{symm-etc}), so the above expression is equivalent to 
\begin{align}
    \frac{1}{2}E=& \frac{1}{2}m \la{v}_x^{2}\ra+\frac{1}{2} m \omega^{2} \la x^{2}\ra \label{E-line1}
    \\
    &+\la X^{2}\ra  \frac{1}{2} \frac{c^{2}}{m \omega^{2}}
    - c \la X x\ra \label{E-line2}
\end{align}

The first line can be found by integration of the equation of motion (\ref{motion-x}):
\begin{equation}
 \frac{1}{2}m \la{v}_x^{2}\ra+\frac{1}{2} m \omega^{2} \la x^{2}\ra \Big\vert_{0}^t=c \int_0^t X(t') v_x(t') \d t'
\end{equation}
Using (\ref{v-sol}) (in terms of $v_x$) we write the first line
\begin{align}
    \label{s68-line1}
    (\ref{E-line1})=&T + \sigma_X \frac{Tc^2}{2 m \omega^2 \omega_0^2} 
    \\
    &+\frac{c^2}{m}\int_0^t \d t'\int_0^{t'} \d t'' \; C_{XX}(t',t'')\cos(\omega(t'-t''))
    \\
    &-\frac{T c^2}{m \omega}\int_0^t \d t'\int_0^{t'} \d t'' \; H_0(t'-t'')\sin(\omega(t'-t''))
    \\
    &-\frac{c^2}{m \omega} \int_0^t \d t' C_{XX}(t',0) \sin(\omega t')
\end{align}
where the line (\ref{s68-line1}) comes from averages at $t=0$. The second line (\ref{E-line2}) can be worked out using (\ref{x-sol})
\begin{align}
    (\ref{E-line2})=& \frac{T c^2}{2m\omega^2\omega_0^2}-C_{XX}(t,0)\frac{c^2}{m\omega^2}\cos(\omega t)
    \\
    &-\frac{c^2}{m\omega}\int_0^t \d t' \; C_{XX}(t,t')\sin(\omega(t-t'))
    \\
    &-\frac{T c^2}{m\omega^2}\int_0^t \d t'\; H_0(t')\cos(\omega t') 
\end{align}
At this stage, all expressions in (\ref{E-line1}, \ref{E-line2}) can be integrated analytically by \textit{symbolic computation software} (c.f. footnote \ref{foot} on page \pageref{foot}). Then we take the limit $t\rightarrow\infty$. So we get
\begin{align}
    E=&
    2T-
    \delta\omega\,\frac{T \gamma}{\pi}
    \left(
    \frac{(1-\sigma_V) \omega^{2}
        +(1-\sigma_X) \omega_0^{2}}{\left(\gamma^{2}+(\omega-b)^{2}\right) \omega^{2}-2(\omega-b) \omega \omega_0^{2}+\omega_0^{4}}
    +\frac{(1-\sigma_V) \omega^{2}
        +(1-\sigma_X)\omega_0^{2}}
        {\left(\gamma^{2}+(\omega+b)^{2}\right)\omega^{2}-2 (b+\omega) \omega\omega_0^{2}+\omega_0^{4}}
        \right)
    \\
    =&2T-\delta\omega\,\frac{T \gamma}{\pi} \frac{\left( \omega^{2}(b^{2}+\gamma^{2})+\left(\omega^{2}-\omega_0^{2}\right)^{2}\right)\left((1-\sigma_V) \omega^{2}+(1-\sigma_X) \omega_0^{2}\right)}
    {\left((\omega^{2}-\omega_0^{2}-b\omega)^2+\gamma^2\omega^2 
    \right)
    \left((\omega^{2}-\omega_0^{2}+b\omega)^2+\gamma^2\omega^2\right)}
\end{align}
where $\sigma_X$ and $\sigma_V$ are defined in (\ref{sigma-def}).

\end{document}